\pdfoutput=1

\documentclass[11pt,twoside,a4paper,cmspaper,final,collab]{cms-tdr}
\def\svnVersion{203645}\def\cmsCernNo{2013-076}\def\cmsCernDate{\today}\def\cmsMessage{Published in Physics Letters B as \href{http://dx.doi.org/10.1016/j.physletb.2013.06.058}{\doi{10.1016/j.physletb.2013.06.058.}}}

\begin{document}\cmsNoteHeader{SUS-12-024}

\hyphenation{had-ron-i-za-tion}
\hyphenation{cal-or-i-me-ter}
\hyphenation{de-vices}

\RCS$Revision: 195607 $
\RCS$HeadURL: svn+ssh://svn.cern.ch/reps/tdr2/papers/SUS-12-024/trunk/SUS-12-024.tex $
\RCS$Id: SUS-12-024.tex 195607 2013-07-11 08:00:12Z bgary $
\newlength\cmsFigWidth
\newlength\cmsFigWidthTwo
\newlength\cmsFigWidthThree
\newlength\cmsFigWidthFigEight
\newlength\cmsFigWidthFigSix
\ifthenelse{\boolean{cms@external}}{\setlength\cmsFigWidth{0.85\columnwidth}}{\setlength\cmsFigWidth{0.4\textwidth}}
\ifthenelse{\boolean{cms@external}}{\setlength\cmsFigWidthTwo{0.85\columnwidth}}{\setlength\cmsFigWidthTwo{0.45\textwidth}}
\ifthenelse{\boolean{cms@external}}{\setlength\cmsFigWidthThree{0.32\textwidth}}{\setlength\cmsFigWidthThree{0.32\textwidth}}
\ifthenelse{\boolean{cms@external}}{\setlength\cmsFigWidthFigEight{0.68\columnwidth}}{\setlength\cmsFigWidthFigEight{0.32\textwidth}}
\ifthenelse{\boolean{cms@external}}{\setlength\cmsFigWidthFigSix{0.95\columnwidth}}{\setlength\cmsFigWidthFigSix{0.8\textwidth}}
\ifthenelse{\boolean{cms@external}}{\providecommand{\cmsLeft}{top}}{\providecommand{\cmsLeft}{left}}
\ifthenelse{\boolean{cms@external}}{\providecommand{\cmsRight}{bottom}}{\providecommand{\cmsRight}{right}}

\providecommand{\Pj}{\mathrm{j}}
\newcommand{\invfb}{\fbinv}
\newcommand{\invpb}{\pbinv}
\newcommand{\met}{\MET}
\newcommand{\nbjet}{\ensuremath{N_{{\cPqb}\text{-jet}}}\xspace}
\newcommand{\dphin}{\ensuremath{\Delta \hat\phi_{\mathrm{min}}}\xspace}
\newcommand{\dphinresi}{\ensuremath{\sigma_{\Delta\phi,i}}\xspace}
\newcommand{\lint}{\ensuremath{\int\lumi\,\rd{t}}\xspace}
\newcommand{\mt}{\ensuremath{m_{\mathrm{T}}}\xspace}
\newcommand{\MHT}{\ensuremath{H_{\mathrm T}^{\text{miss}}}\xspace}
\newcommand{\wpjets}{\PW+jets\xspace}
\newcommand{\zmumu }{\ensuremath{{\Z}\rightarrow\Pgmp\Pgmm}\xspace}
\newcommand{\znunu }{\ensuremath{{\Z}\rightarrow \nu \bar{\nu}}\xspace}
\newcommand{\zee }{\ensuremath{{\Z}\rightarrow\Pep\Pem}\xspace}
\newcommand{\zll }{\ensuremath{{\Z}\rightarrow \ell^+ \ell^-}\xspace}
\newcommand{\vsmtvs }{\rule[-0.3cm]{0cm}{0.75cm}}
\newcommand{\smtvs }{\rule[-0.3cm]{0cm}{0.9cm}}
\newcommand{\tonebbbb}{T1bbbb\xspace}
\newcommand{\tonetttt}{T1tttt\xspace}
\newcommand{\mlsp}{\ensuremath{m_{{\PSGcz}_1}}\xspace}
\newcommand{\mgluino}{\ensuremath{m_{\sGlu}}\xspace}
\newcommand{\FEWZ} {{\textsc{fewz}}\xspace}

\cmsNoteHeader{AN-12-081} 
\title{Search for gluino mediated bottom- and top-squark production in multijet final states in  pp collisions at 8\TeV}

\date{\today}

\abstract{
A search for supersymmetry is presented based on events
with large missing transverse energy,
no isolated electron or muon,
and at least three jets
with one or more identified as a bottom-quark jet.
A simultaneous examination is performed of the
numbers of events in exclusive bins of
the scalar sum of jet transverse momentum values,
missing transverse energy,
and bottom-quark jet multiplicity.
The sample,
corresponding to an integrated luminosity of 19.4\fbinv,
consists of proton-proton collision data recorded at a center-of-mass energy of 8\TeV
with the CMS detector at the LHC in 2012.
The observed numbers of events are found to be consistent with the
standard model expectation,
which is evaluated with control samples in data.
The results are interpreted in the context of two simplified
supersymmetric scenarios in which gluino pair production is followed
by the decay of each gluino to an undetected lightest supersymmetric particle
and either a bottom or top quark-antiquark pair,
characteristic of gluino mediated bottom- or top-squark production.
Using the production cross section calculated to
next-to-leading-order plus next-to-leading-logarithm accuracy,
and in the limit of a massless lightest supersymmetric particle,
we exclude gluinos with masses below 1170\GeV
and 1020\GeV for the two scenarios, respectively.
}

\hypersetup{%
pdfauthor={CMS Collaboration},%
pdftitle={Search for gluino mediated bottom- and top-squark production in multijet final states in  pp collisions at 8 TeV},%
pdfsubject={CMS},%
pdfkeywords={CMS, physics, SUSY, hadronic, btag}}

\maketitle 

\section{Introduction}
\label{sec-introduction}

The standard model (SM) of particle physics has proved
to be remarkably successful in describing phenomena
up to the highest energy scales that have been probed.
Nonetheless,
the SM is widely viewed to be incomplete.
Many extensions have been proposed to provide a more fundamental theory.
Supersymmetry (SUSY)~\cite{Ramond:1971gb,Golfand:1971iw,Neveu:1971rx,
Volkov:1972jx,Wess:1973kz,Wess:1974tw,Fayet:1974pd,Nilles:1983ge},
one such extension,
postulates that each SM particle is paired
with a SUSY partner from which it differs in spin by one-half unit,
with otherwise identical quantum numbers.
For example,
squarks and gluinos are the SUSY partners of quarks and gluons,
respectively.
One of the principal motivations for SUSY is to stabilize the calculation of
the Higgs boson mass.
For this stabilization to be
``natural''~\cite{Dimopoulos:1995mi,Barbieri:2009ev,Papucci:2011wy},
top squarks, bottom squarks, and to a lesser extent gluinos,
must be relatively light.
If top and bottom squarks are light,
their production is enhanced,
either through direct pair production or through production mediated by gluinos,
where the latter process is favored
if the gluino is relatively light so that its pair production cross section is large.
Since the decay products of both bottom and top squarks include bottom quarks,
natural SUSY models are characterized by
an abundance of bottom-quark jets ({\cPqb} jets).

In R-parity-conserving~\cite{bib-rparity} SUSY models,
supersymmetric particles are created in pairs.
Each member of the pair initiates a decay chain
that terminates with the lightest SUSY particle (LSP) and SM particles,
typically including quarks and gluons,
which then hadronize to form jets.
If the LSP only interacts weakly,
as in the case of a dark-matter candidate,
it escapes detection,
potentially yielding significant missing transverse energy (\met).
Thus large values of \met provide another possible SUSY signature.

In this Letter,
we present a search for SUSY in events with at least three jets,
one or more of which are identified as \cPqb\ jets (\cPqb\ tagged),
and large~\met.
The search is based on a sample of proton-proton (pp) collision data collected
at $\sqrt{s}=8$\TeV with the Compact Muon Solenoid (CMS) detector at the
CERN Large Hadron Collider (LHC) in 2012,
corresponding to an integrated luminosity of 19.4\invfb.
Previous LHC new-physics searches in final states with {\cPqb} jets and \met are presented
in Refs.~\cite{Aad:2011ks,Aad:2011cw,ATLAS:2012ah,Aad:2012pq,Aad:2012si,Aad:2012ar,Aad:2012yr,
Chatrchyan:2011bj,Chatrchyan:2012sa,Chatrchyan:2012jx,RA2b2011pub,Chatrchyan:2012wa,
Chatrchyan:2012pca}.
The current analysis is an extension of the study presented in Ref.~\cite{RA2b2011pub},
which was based on 4.98\invfb of data collected at $\sqrt{s}=7$\TeV.
We retain the same basic analysis procedures,
characterized by a strong reliance on control samples in data,
to evaluate the SM backgrounds.
The principal backgrounds arise from the production of
events with a top quark-antiquark (\ttbar) pair,
a single-top quark,
a $\PW$ boson in association with jets ($\PW$+jets),
a {\Z} boson in association with jets ({\Z}+jets),
and multiple jets produced through the strong interaction,
in which a b-tagged jet is present.
We refer to events in the latter category as
quantum chromodynamics (QCD) multijet events.
For {\PW}+jets events and events with top quarks,
significant \met can arise if a {\PW} boson decays into
a neutrino and a charged lepton.
The neutrino provides a source of genuine \met.
For events with a {\Z} boson,
significant \met can arise if the {\Z} boson decays to two neutrinos.
For QCD multijet events,
significant \met can arise
when a charm or bottom quark undergoes semileptonic decay,
but the main source of \met is a mismeasurement of
jet transverse momentum~\pt.
The QCD multijet category excludes events that are contained in the other categories.

\begin{figure}[bt]
\centering
\includegraphics[width=\cmsFigWidthTwo]{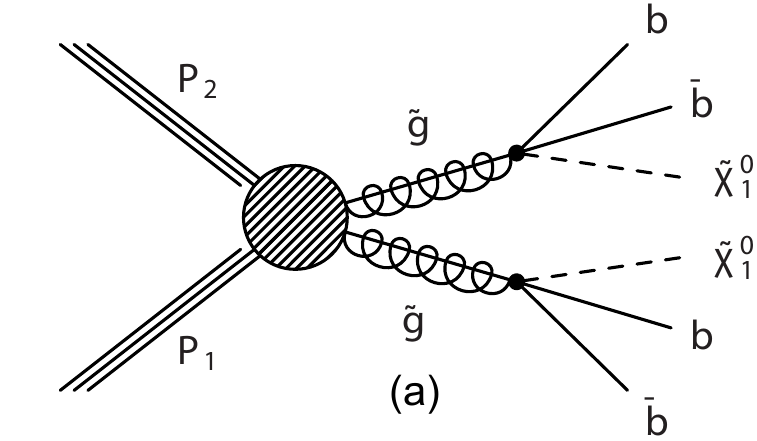}
\includegraphics[width=\cmsFigWidthTwo]{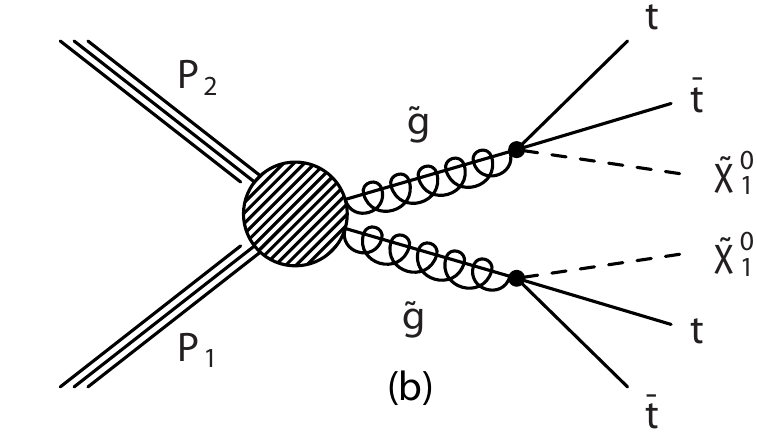}
\caption{
Event diagrams for the (a) \tonebbbb and (b)~\tonetttt
simplified SUSY scenarios.
}
\label{fig:T1bbbb-diagram}
\end{figure}

As new-physics scenarios,
we consider the simplified SUSY spectra~\cite{bib-sms-1,bib-sms-2,bib-sms-3,bib-sms-4}
in which gluino pair production is followed by the
decay of each gluino~\sGlu into a bottom quark and an off-shell bottom squark
or into a top quark and an off-shell top squark.
The off-shell bottom (top) squark decays into a bottom
(top) quark and the LSP,
where the LSP is assumed to escape detection,
leading to significant~\met.
A possible LSP candidate is the lightest neutralino~${\PSGcz}_1$;
we therefore use the symbol ${\PSGcz}_1$ to denote the LSP.
We assume all SUSY particles other than the gluino
and the LSP to be too heavy to be produced at current LHC energies,
and the gluino to be short-lived.
The production cross section is
computed~\cite{bib-nlo-nll-01,bib-nlo-nll-02,bib-nlo-nll-03,bib-nlo-nll-04,bib-nlo-nll-05}
at the next-to-leading order (NLO) plus next-to-leading-logarithm (NLL) level.
We denote the $\sGlu\sGlu\rightarrow 2\times\cPqb\overline{\cPqb}{\PSGcz}_1$
process as the \tonebbbb scenario
and the $\sGlu\sGlu\rightarrow2\times\cPqt\overline{\cPqt}{\PSGcz}_1$
process as the \tonetttt scenario~\cite{Chatrchyan:2013wc}.
Event diagrams are shown in Fig.~\ref{fig:T1bbbb-diagram}.
If the bottom (top) squark is much lighter than any other squark
under the conditions described above,
gluino decays are expected to be dominated by the three-body process
of Fig.~\ref{fig:T1bbbb-diagram}a (\ref{fig:T1bbbb-diagram}b).
The gluino and LSP masses are treated as independent parameters.

\begin{figure*}[tb]
\centering
\includegraphics[width=0.95\linewidth]{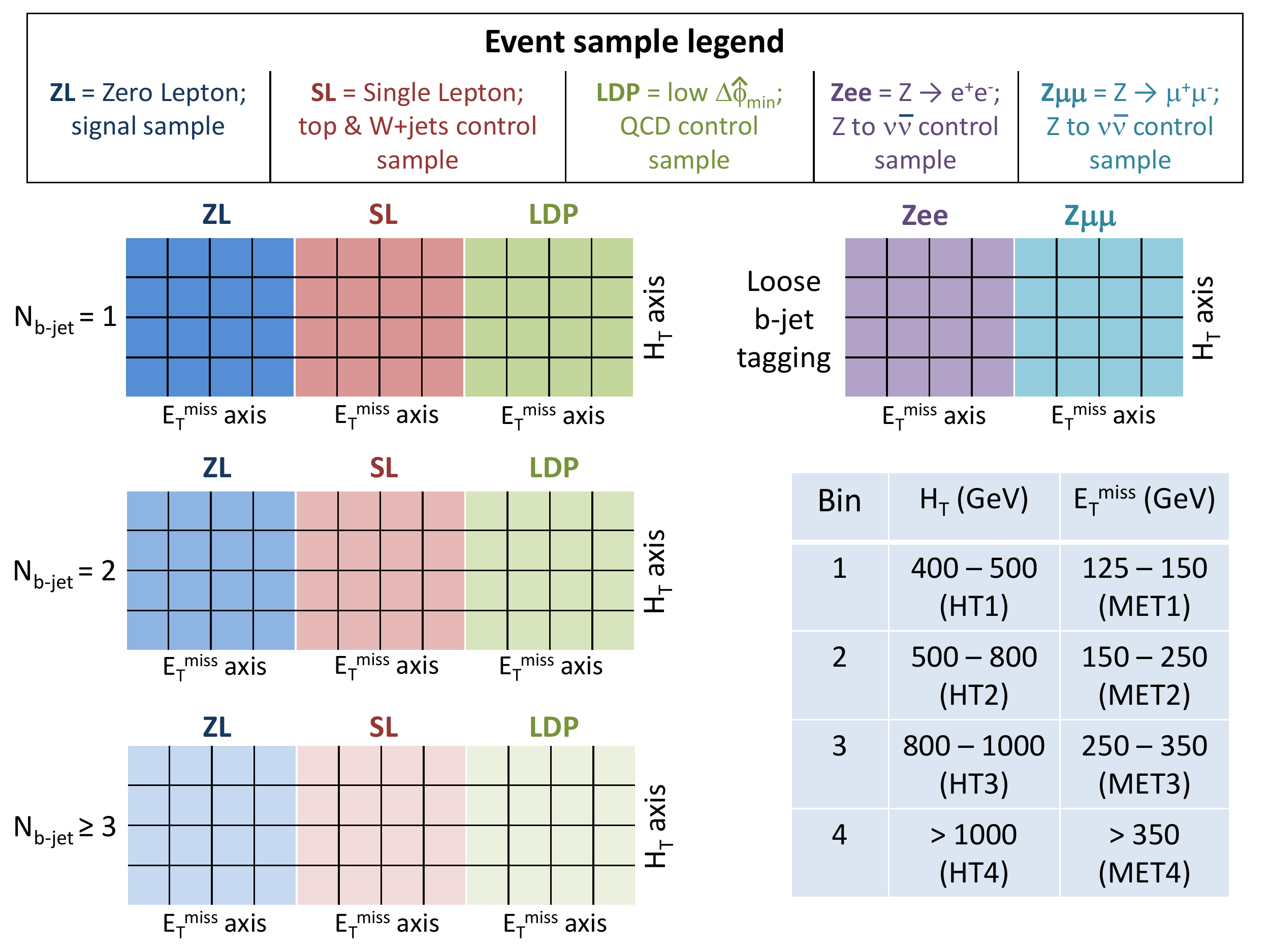}
\caption{
Schematic diagram illustrating the 176 mutually exclusive bins
in the analysis.
The \met and \HT distributions are divided into four bins each;
the table gives the bin definitions.
The designations HT$i$ and MET$i$ ($i=1$--4)
are used to label the individual \HT and \MET bins.
The \nbjet
distributions of the signal sample (ZL),
top-quark and {\PW}+jets control sample (SL),
and QCD multijet control sample (LDP),
contain three bins each,
corresponding to exactly one,
exactly two,
and three or more identified {\cPqb} jets.
}
\label{fig:binning-cartoon}
\end{figure*}

It is rare for a \tonebbbb event to contain an isolated high-\pt lepton.
To define the search region for this study,
we therefore veto events
with an identified isolated electron or muon.
We also veto events with an isolated charged track,
characteristic of $\tau$-lepton decay.
The resulting collection of events is referred to as the
zero-lepton (ZL) or ``signal'' sample.
Besides the ZL sample,
control samples are defined in order to evaluate the SM background.
To evaluate the backgrounds from top-quark and {\PW}+jets events
(where ``top quark'' refers to both \ttbar and single-top-quark events),
we select a top-quark- and {\PW}+jets-dominated
control sample by requiring the presence of
exactly one identified isolated electron or muon.
We refer to this sample as the single-lepton (SL) sample.
(Top-quark and {\PW}+jets events are grouped into a
single background category because of their similar experimental signatures.)
To evaluate the QCD multijet background,
we employ the minimum normalized
azimuthal angle \dphin~\cite{RA2b2011pub}
between the \met vector and one of the three highest-\pt jets,
selecting a QCD-dominated control sample by requiring
small values of this variable.\footnote{For the current study,
we use a slightly modified definition of the \dphin variable compared
to Ref.~\cite{RA2b2011pub}:
we now use ``$\arcsin$'' rather than ``$\arctan$''
in the expression for \dphinresi
(see Section~IV of Ref.~\cite{RA2b2011pub}).
This modification introduces a negligible difference for
the small angles relevant here.
Nonetheless, the modified expression is technically more correct
than the original one.
}
We refer to this control sample as the low-\dphin (LDP) sample.
The {\Z}+jets background is evaluated with control
samples of \zll\ events ($\ell=\Pe$ and {\Pgm}).
Our analysis is performed in the framework of a global likelihood fit that
simultaneously analyzes the signal and background content,
accounting for signal contributions to the ZL and control samples
in a unified and consistent manner.

In contrast to \tonebbbb events,
events in the \tonetttt scenario are expected to appear
in both the ZL and SL samples.
Since our global likelihood fitting procedure can account for
\tonetttt contributions to the control samples,
the analysis procedures and background evaluation methods used
to examine the \tonetttt scenario are essentially the same as
those used for the \tonebbbb scenario.

This study extends the analysis of Ref.~\cite{RA2b2011pub}
by exploiting the expected differences in shape between the
\tonebbbb or \tonetttt scenario
and each of the SM background components in the distributions of
\met, the number \nbjet of {\cPqb}-tagged jets in an event,
and \HT,
where \HT is the scalar sum of jet \pt values.
(The quantitative definitions of \met and \HT are given in Section~\ref{sec:basiccuts}.)
The data are divided into mutually exclusive bins in these three variables,
as indicated schematically in Fig.~\ref{fig:binning-cartoon}.
The \met and \HT distributions are divided into four bins each.
The definitions of these bins are given in the table
of Fig.~\ref{fig:binning-cartoon}.
For the ZL, SL, and LDP samples,
the {\cPqb}-jet multiplicity distribution is divided into three bins,
corresponding to $\nbjet=1$, 2, or $\geq 3$.
There are 176 mutually exclusive bins of data in the analysis,
48 each for the ZL, SL, and LDP samples,
and 16 each for the {\zee} and {\zmumu} samples.
The contents of the bins are examined simultaneously
in the likelihood fit.

This Letter is organized as follows.
In Section~\ref{sec-detector} we discuss the detector and trigger.
Sections~\ref{sec:basiccuts} and~\ref{sec:searchregions}
describe the event selection.
The likelihood framework and background determination methods
are presented in Section~\ref{sec:likelihood}.
Section~\ref{sec:results} presents the results
and Section~\ref{sec:summary} a summary.

\section{Detector and trigger}
\label{sec-detector}

A detailed description of the CMS detector is
given elsewhere~\cite{Chatrchyan:2008aa}.
The CMS coordinate system is defined with the origin at the center
of the detector and the $z$ axis along the direction of the
counterclockwise beam.
The transverse plane is perpendicular to the beam axis,
with $\phi$ the azimuthal angle (measured in radians),
$\theta$ the polar angle,
and $\eta=-\ln[\tan(\theta/2)]$ the pseudorapidity.
A superconducting solenoid
provides an axial magnetic field of 3.8\unit{T}.
Within the field volume are a silicon pixel and strip tracker,
a crystal electromagnetic calorimeter,
and a brass-scintillator hadron calorimeter.
The tracking system is completed with muon detectors,
based on gas-ionization chambers embedded in the
steel flux-return yoke outside the solenoid.
The tracking system covers $\abs{\eta}<2.5$
and the calorimeters $\abs{\eta}<3.0$.
The $3<\abs{\eta}<5$ region is instrumented with forward calorimeters.
The near-hermeticity of the detector permits accurate
measurements of energy balance in the transverse plane.

Events are selected using multiple trigger conditions,
based primarily on thresholds for \HT and \met.
The trigger efficiency,
determined from data,
is the probability for a signal or control sample
event to satisfy the trigger conditions.
In our analysis,
the data are examined in exclusive regions of \HT and \met,
as described above.
The trigger is found to be nearly 100\% efficient
except in regions with low values of both \HT and \met.
In the bin with lowest \HT and \met,
i.e., the HT1-MET1 bin of Fig.~\ref{fig:binning-cartoon},
the evaluated trigger efficiency is
$0.91\pm0.01$ ($0.86\pm0.09$)
for the trigger relevant for the ZL and SL (LDP) samples.
Corrections are applied to account for the trigger efficiencies
and their corresponding uncertainties.

\section{Event selection}
\label{sec:basiccuts}

Physics objects are defined with the particle flow
(PF) method~\cite{bib-cms-pf},
which is used to reconstruct and identify charged and neutral hadrons,
electrons (with associated bremsstrahlung photons), muons, and photons,
using an optimized combination of information from CMS subdetectors.
Tau leptons are identified using the reconstructed PF objects.
The event primary vertex is identified by selecting the reconstructed vertex
that has the largest sum of charged-track $\pt^2$ values.
Events are required to have a primary vertex with
at least four charged tracks
and that lies within 24\cm of the origin in the direction along the beam axis
and 2\cm in the perpendicular direction.
Charged particles used in the analysis must
emanate from the primary vertex.
In this way,
charged particles associated with extraneous pp interactions
within the same bunch crossing (``pileup'') are disregarded.
The PF objects serve as input for jet reconstruction,
based on the anti-\kt algorithm~\cite{bib-antikt}
with a distance parameter of~0.5.
Jet corrections
are applied in both \pt and~$\eta$ to account for
residual effects of non-uniform detector response.
Additional corrections~\cite{Chatrchyan:2011ds,Cacciari:2011ma}
account for pileup effects from neutral particles.
The missing transverse energy \MET is defined as the modulus
of the vector sum of the transverse momenta of all PF objects.
The \MET vector is the negative of that same vector sum.

The requirements used to select the zero-lepton (ZL) event sample
are as follows:
\begin{itemize}
\item at least three jets with $\pt>50$\GeV and $\abs{\eta}<2.4$,
   where the two leading jets satisfy $\pt>70\GeV$;
\item $\HT > 400 \GeV$, where \HT is calculated using jets with
    $\pt>50$\GeV and $\abs{\eta} < 2.4$;
\item $\met > 125 \GeV$;
\item no identified, isolated electron or muon candidate with $\pt>10\GeV$;
  electron candidates are restricted to $\abs{\eta}<2.5$ and
  muon candidates to $\abs{\eta}<2.4$;
\item no isolated charged-particle track with $\pt>15\GeV$ and $\abs{\eta}<2.4$;
\item $\dphin> 4.0$, where the \dphin variable
  is described in Ref.~\cite{RA2b2011pub};
\item at least one \cPqb-tagged jet,
  where \cPqb-tagged jets are required to have $\pt>50$\GeV and $\abs{\eta}<2.4$.
\end{itemize}
The isolated-track requirement eliminates events with
an isolated electron or muon
in cases where the lepton is not identified,
as well as events with a $\tau$ lepton that decays hadronically.
Electrons, muons, and tracks are considered isolated if the scalar sum of the
\pt values of charged hadrons
(for electrons and muons, also photons and neutral hadrons)
surrounding the lepton or track within a cone of radius
$\sqrt{(\Delta\eta)^2+(\Delta\phi)^2}=0.3$ (0.4 for muons),
divided by the lepton or track \pt value itself,
is less than 0.15, 0.20, and 0.05, respectively.

Identification of {\cPqb} jets is based on
the combined-secondary-vertex algorithm described in Ref.~\cite{Chatrchyan:2012jua}
(we use the ``medium'' working point).
This algorithm combines information about secondary vertices,
track impact parameters, and jet kinematics,
to separate {\cPqb} jets from
light-flavored-quark, charm-quark, and gluon jets.
The nominal {\cPqb}-tagging efficiency is 75\%
for jets with a \pt value of 80\GeV,
as determined from a sample of simulated
{\cPqb}-jet-enriched events~\cite{Chatrchyan:2012jua}.
The corresponding misidentification rate for light-quark jets is~1.0\%.

\section{Control samples, search regions, and event simulation}
\label{sec:searchregions}

The top-quark- and {\wpjets}-dominated SL control sample is defined by selecting
events with exactly one electron or one muon,
using the lepton selection criteria and all other nominal selection requirements
given in Section~\ref{sec:basiccuts},
with the exception of the requirement that there be no isolated track.
To reduce potential contributions from signal \tonetttt events,
we apply an additional requirement $\mt<100\GeV$ to the SL sample only,
where
$\mt=\{2\met\pt^\ell[1-\cos(\Delta\phi_{\ell,\met})]\}^{1/2}$
is the transverse mass formed from the
\met and $\pt^\ell$ (lepton transverse momentum) vectors,
with $\Delta\phi_{\ell,\met}$ the corresponding difference in the azimuthal angle.

The region $\dphin<4$,
with all other nominal selection requirements from
Section~\ref{sec:basiccuts} imposed,
defines the QCD-dominated LDP control region.

To evaluate the {\Z}+jets background,
we select {\Z}+jets control samples with {\zee} and {\zmumu} decays,
as described in Section~\ref{sec:like-znunu}.

The data are divided into mutually exclusive bins of \met,
\HT, and \nbjet,
as shown in Fig.~\ref{fig:binning-cartoon}.
This binning is chosen based on simulation studies with
SUSY signal and SM background event samples,
for which signal sensitivity in the presence of SUSY events,
and limits in the absence of such events,
are both considered.
The best performance is obtained with relatively narrow bins
at low \HT\ and \met,
which help to characterize the background shapes,
and with multiple bins at high \HT\ and \met,
which provide regions with reasonable signal efficiency
and very little background.
Within this general framework,
the sensitivity is found to be relatively independent of particular
binning choices.

\begin{figure*}[tbp]
\begin{center}
\includegraphics[width=\cmsFigWidthThree]{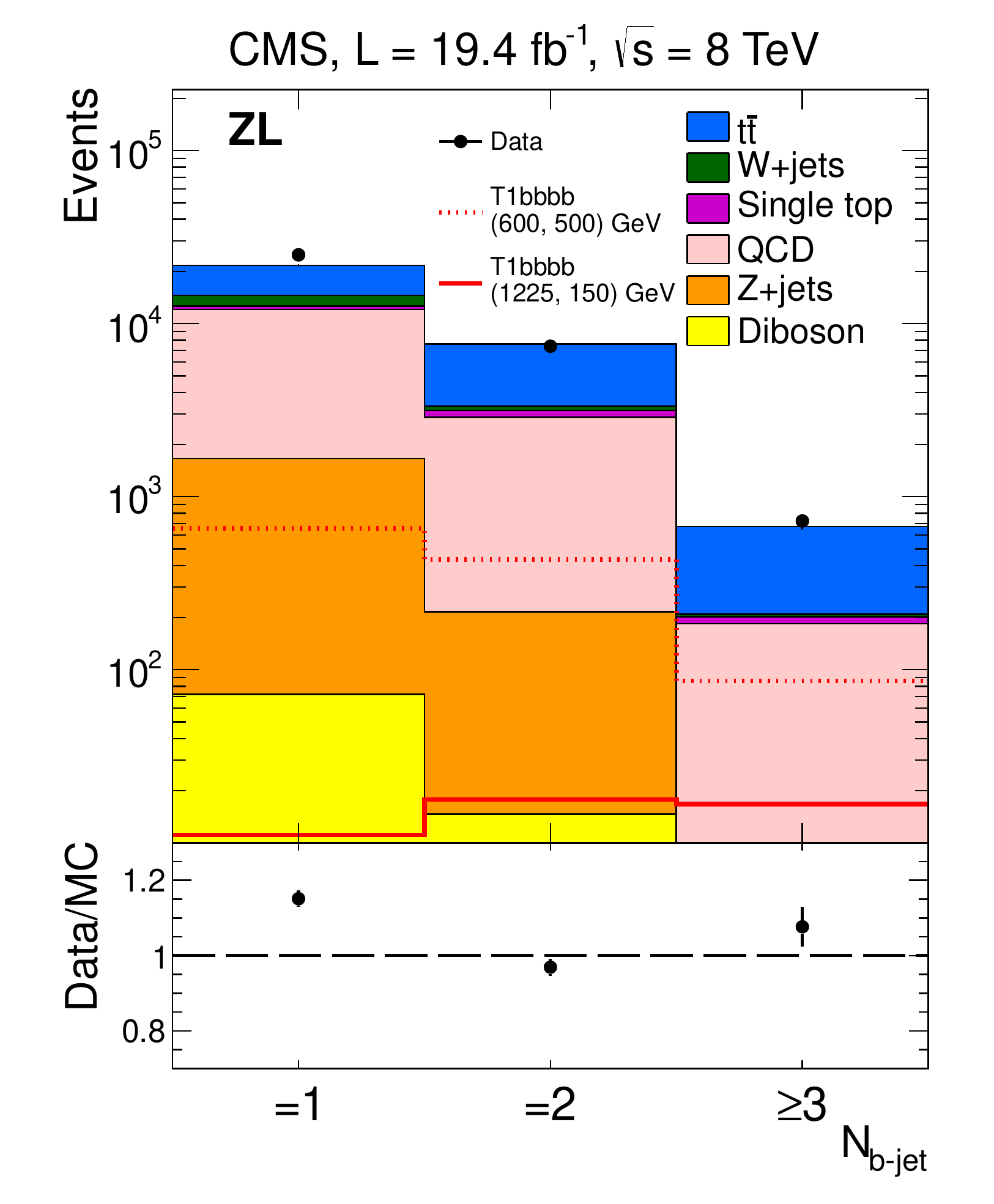}
\includegraphics[width=\cmsFigWidthThree]{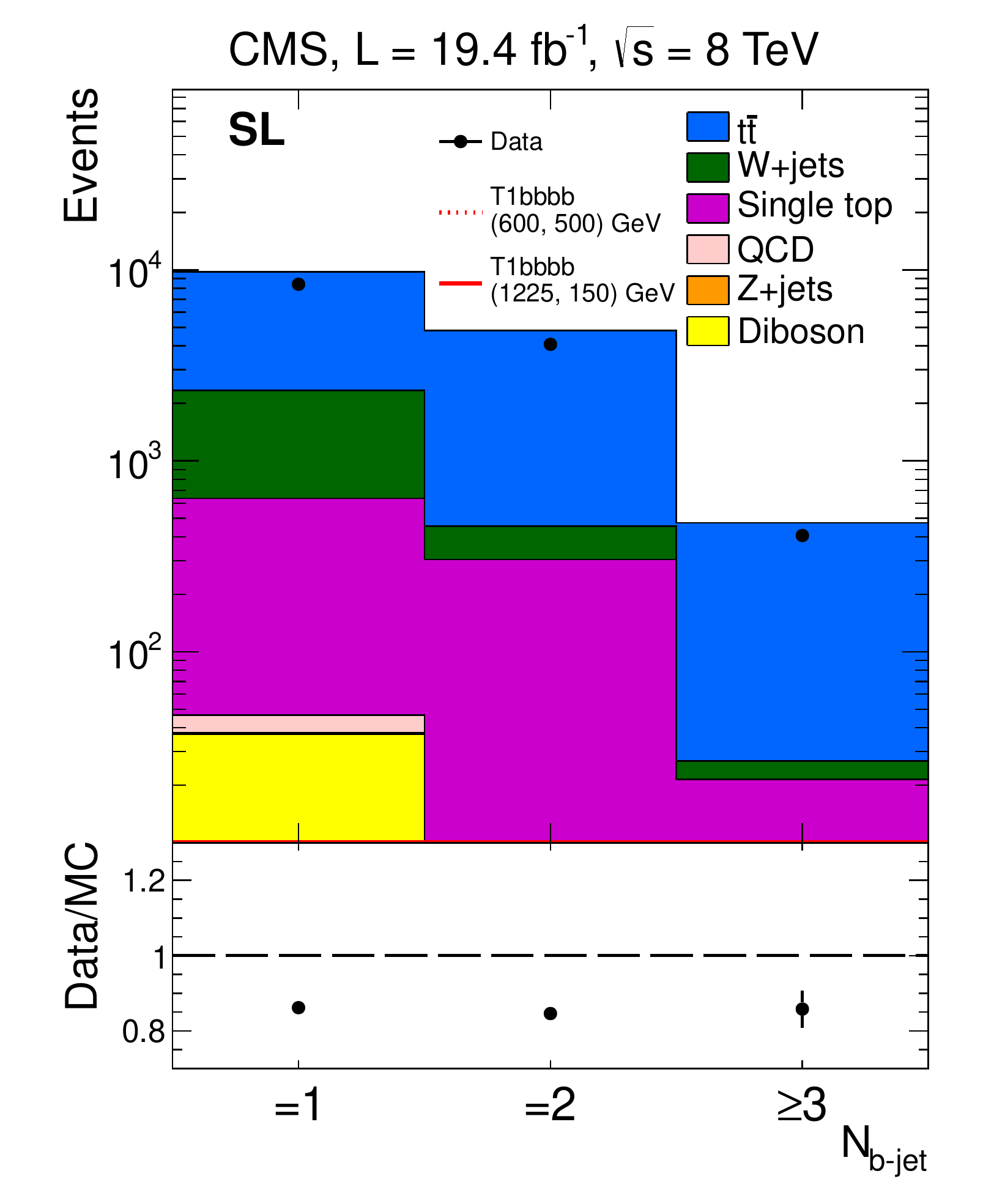}
\includegraphics[width=\cmsFigWidthThree]{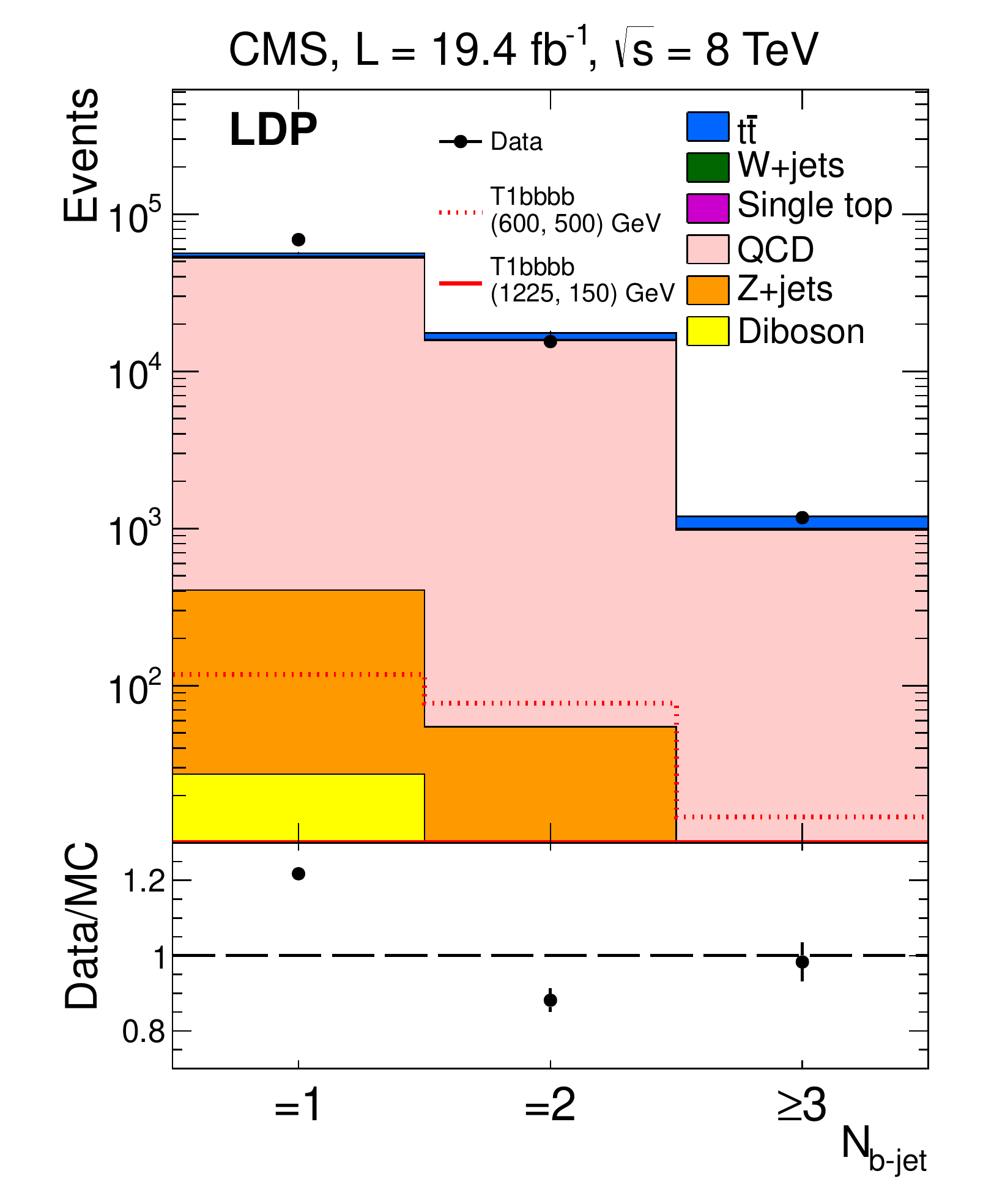}
\includegraphics[width=\cmsFigWidthThree]{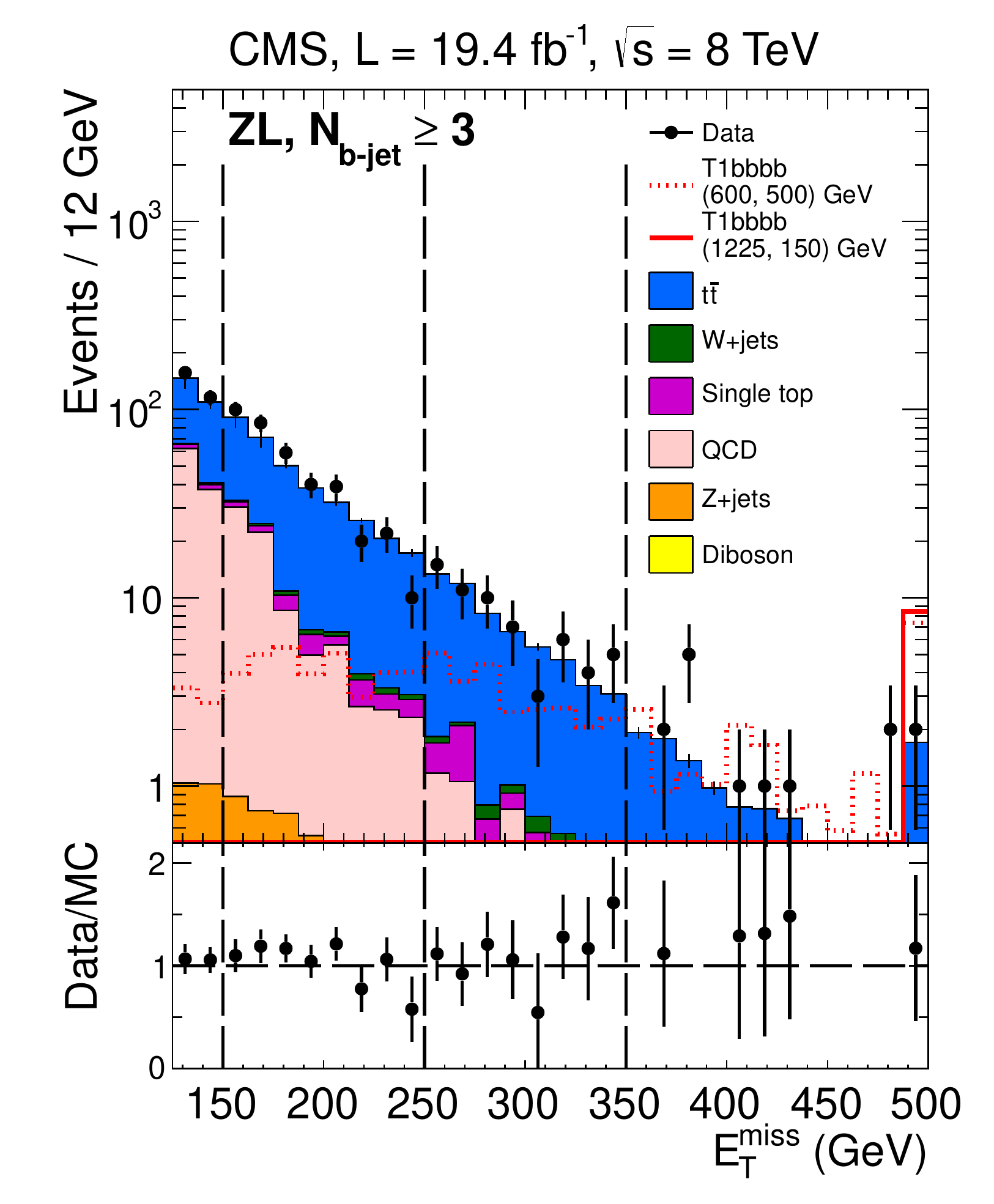}
\includegraphics[width=\cmsFigWidthThree]{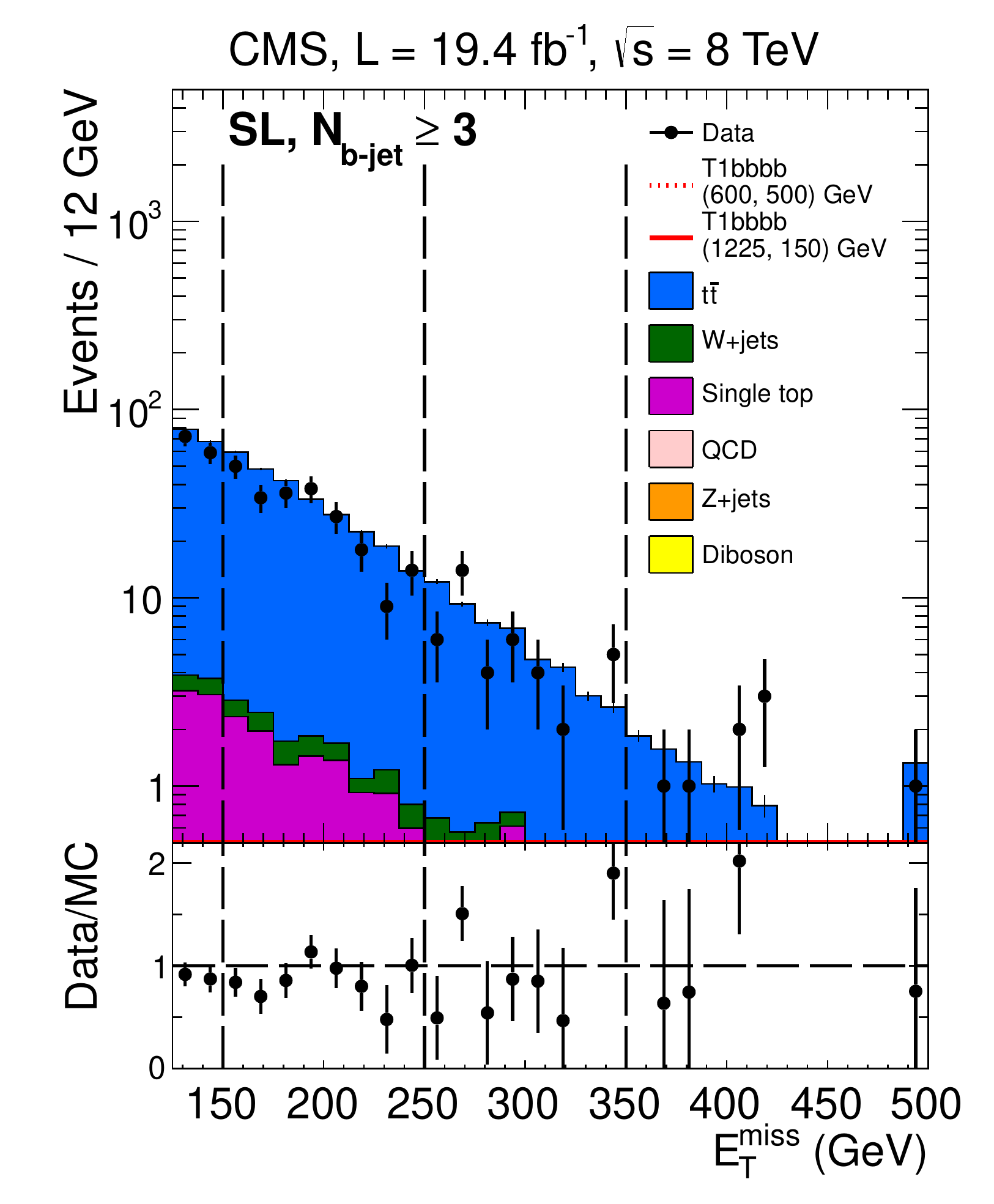}
\includegraphics[width=\cmsFigWidthThree]{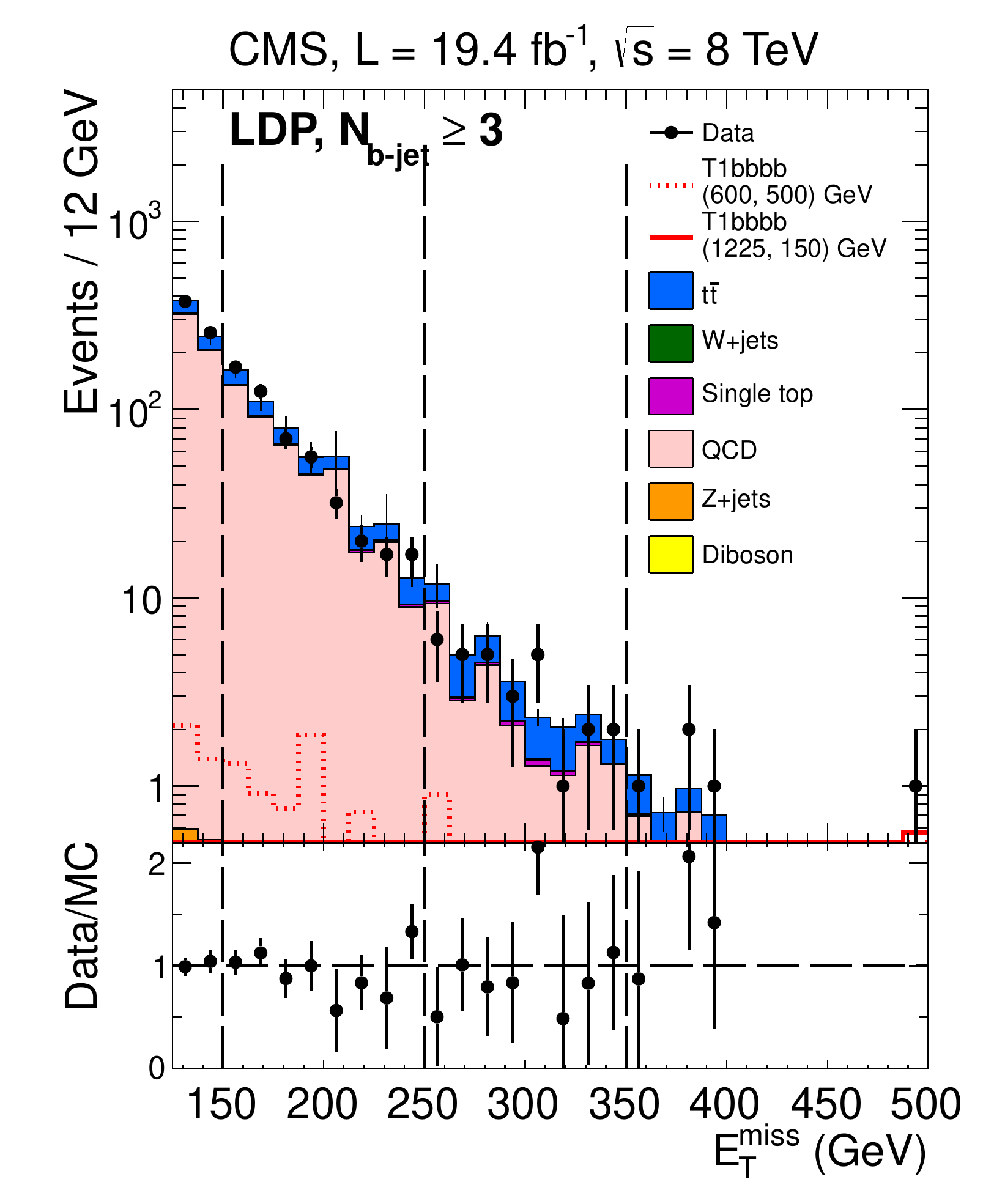}
\includegraphics[width=\cmsFigWidthThree]{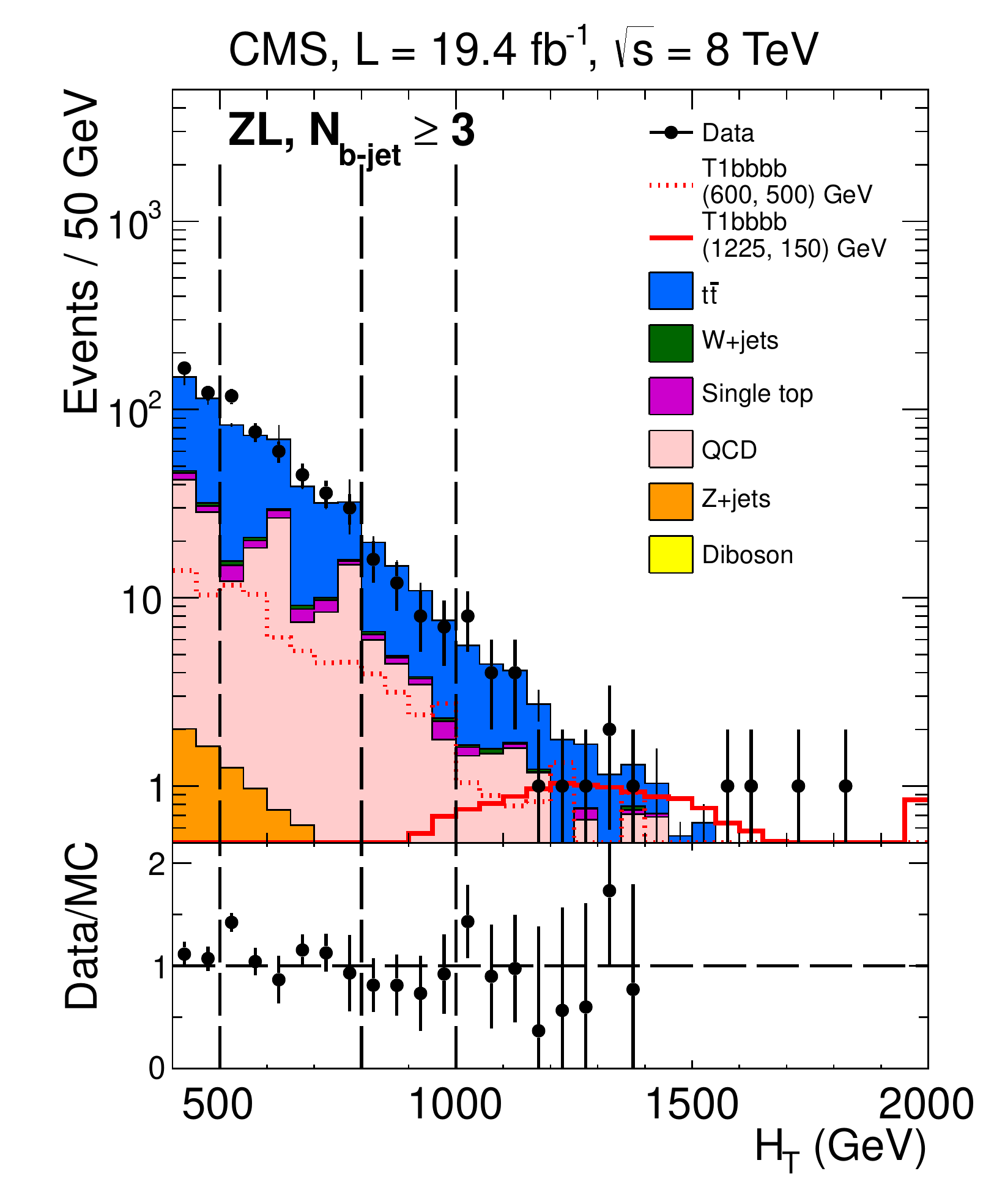}
\includegraphics[width=\cmsFigWidthThree]{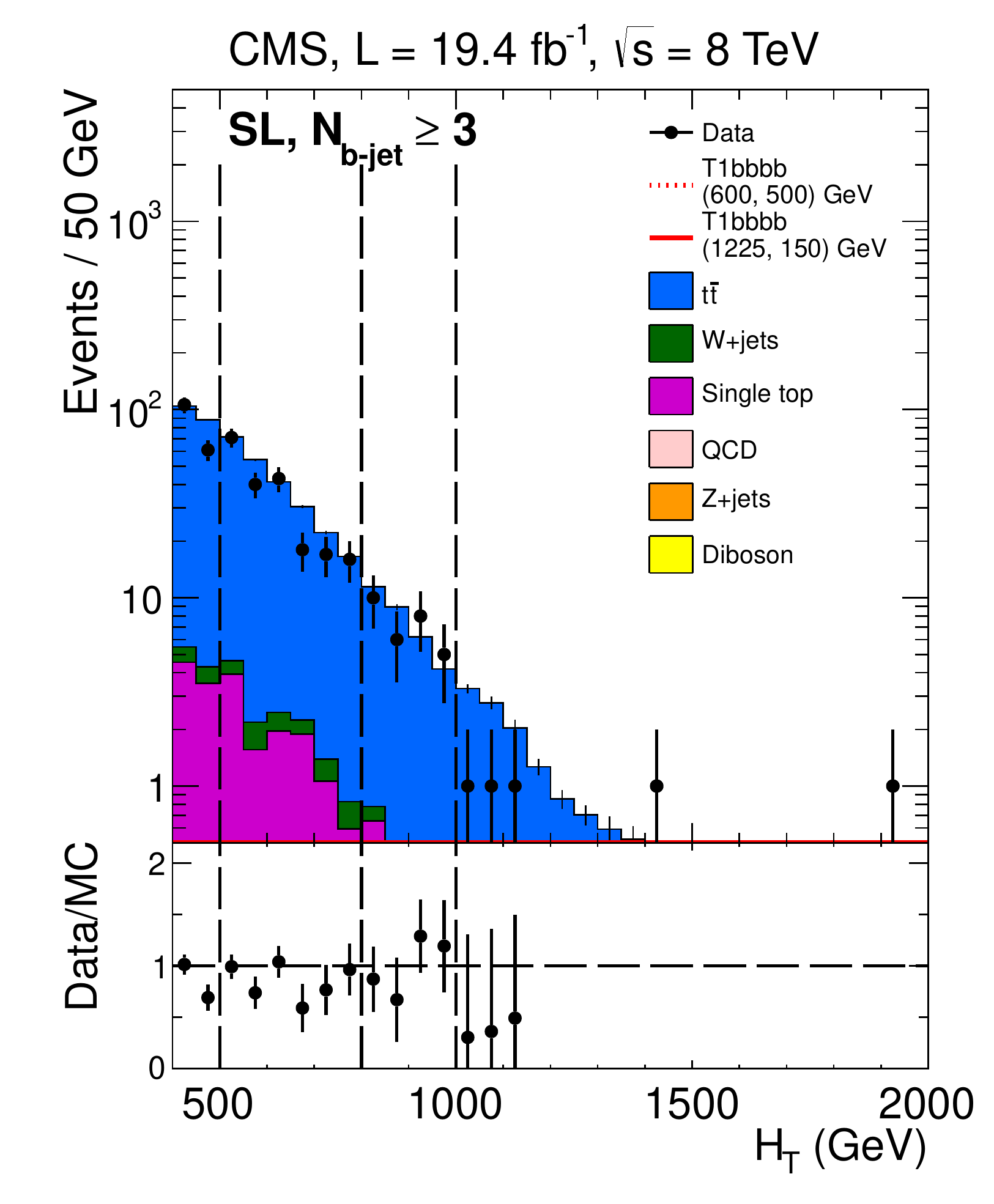}
\includegraphics[width=\cmsFigWidthThree]{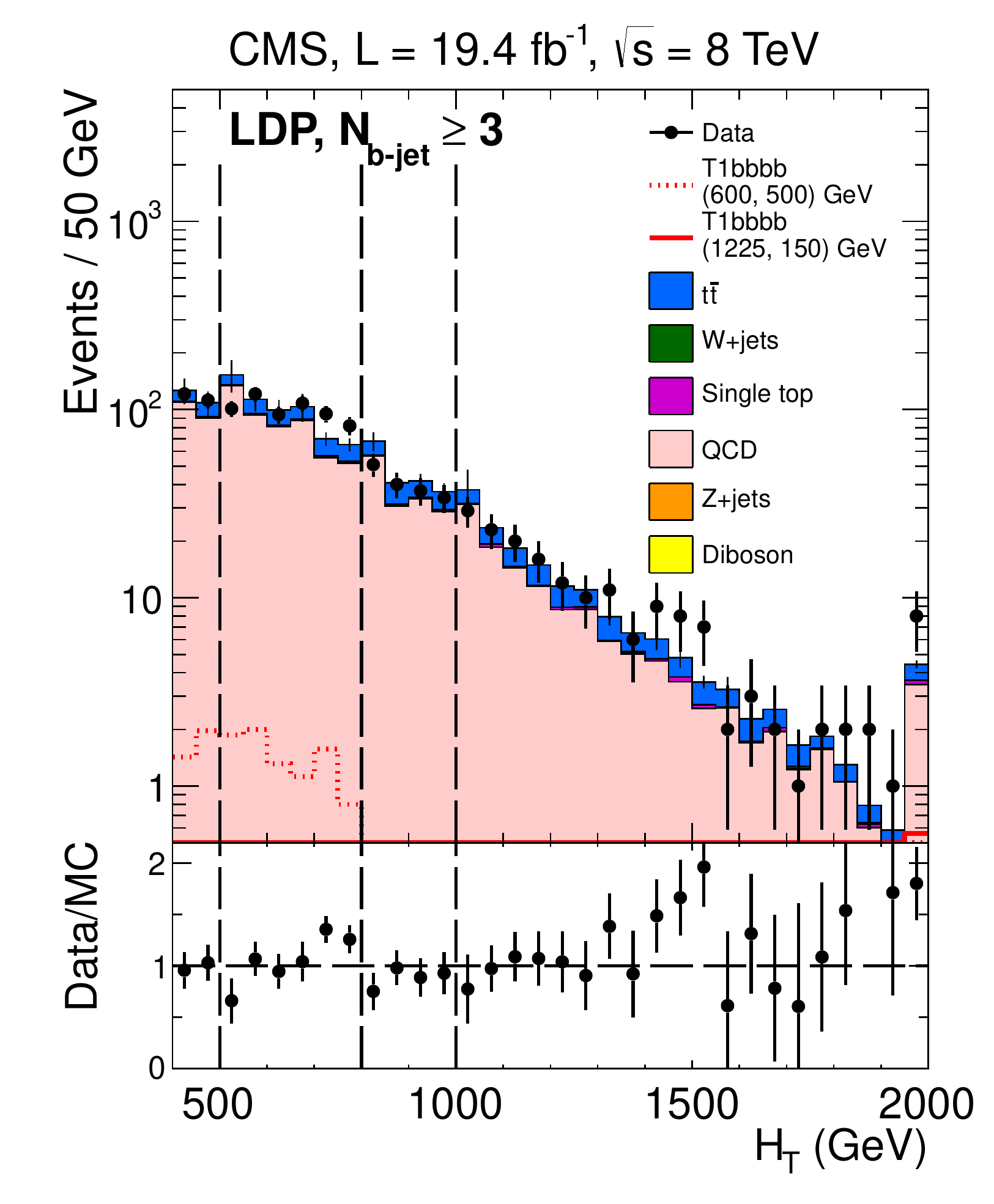}
\caption{
[top row]
Data and Monte Carlo distributions of the number \nbjet of \cPqb-tagged jets
for the [left column] signal (ZL) sample,
[center column] top-quark and {\PW}+jets (SL) control sample,
and [right column] QCD multijet (LDP) control sample.
The lower panes show the ratio of the measured to the simulated events.
[center row] Corresponding \met distributions,
and [bottom row] \HT distributions,
for events with \mbox{$\nbjet\geq 3$.}
The dashed vertical lines indicate the divisions between
the four bins of~\met or~\HT.
Results for the \tonebbbb scenario with
$(\mgluino,\mlsp)=(600\GeV, 500\GeV)$ and $(1225\GeV, 150\GeV)$
are shown as unstacked distributions.
For all results,
the uncertainties are statistical only.
The normalization of the simulated curves is based on the absolute cross sections,
as described in the text.
}
\label{fig:data-mc-compare}
\end{center}
\end{figure*}

To illustrate the characteristics of the events,
Fig.~\ref{fig:data-mc-compare} presents the distribution of
\nbjet for the signal (ZL) and control-region (SL, LDP) samples,
and the corresponding distributions of \met and \HT for $\nbjet\geq 3$.
The results are shown in comparison with Monte Carlo (MC)
simulations of SM processes.
The \ttbar, {\PW}+jets, and {\Z}+jets MC samples are simulated
at the parton level with the {\MADGRAPH} {5.1.1.0}~\cite{bib-madgraph} event generator.
Single-top-quark events are produced with the \POWHEG 301~\cite{bib-powheg} program.
The \PYTHIA 6.4.22~\cite{bib-pythia} generator is used for diboson and QCD multijet events.
For all SM MC samples,
the \GEANTfour~\cite{bib-geant} package is used to model the detector.
The top-quark MC distributions are normalized to an approximate
next-to-next-to-leading order (NNLO) cross-section
calculation~\cite{Kidonakis:2011wy,Kidonakis:2012db}
and the simulated {\PW}+jets and {\Z}+jets
results to the inclusive NNLO cross sections from the
{\FEWZ} generator~\cite{Gavin:2010az}.
The diboson MC distribution,
given by the sum of contributions from {\PW\PW}, {\PW\Z}, and {\Z}{\Z} events,
is normalized to NLO using the cross section from the
{\MCFM} generator~\cite{Campbell:2011bn}.
The QCD multijet distribution is normalized to leading order.
We also consider Drell--Yan events,
generated with {\MADGRAPH} and normalized to NNLO~\cite{Gavin:2010az}.
The contribution of Drell-Yan events is found to be small
(at most one fifth the contribution of diboson
events in all signal regions)
and is not included in Fig.~\ref{fig:data-mc-compare}.

In general,
the simulation is seen to agree with the data,
although some features exhibit differences on the order of 20\%.
Note that these MC results are not used in the analysis
but merely provide guidance on the expected background composition.

Signal \tonebbbb and \tonetttt MC samples are generated for
a range of gluino \mgluino and LSP \mlsp mass values,
with $\mlsp<\mgluino$.
The signal samples are based on \MADGRAPH,
with up to two partons present in addition to the gluino pair.
The decays of the gluino are described using a pure phase-space matrix element in \PYTHIA.
To reduce computational requirements,
the detector is modeled with the CMS fast simulation
program~\cite{Orbaker:2010zz,bib-cms-fastsim-02},
with corrections
to account for modest differences observed with respect to the \GEANTfour simulation.
Figure~\ref{fig:data-mc-compare} includes the distributions of
two representative \tonebbbb scenarios,
one with $(\mgluino,\mlsp)=(600\GeV, 500\GeV)$ and the other with
$(\mgluino,\mlsp)=(1225\GeV, 150\GeV)$,
both of which are at the limit of our expected sensitivity
(Section~\ref{sec:results}).

All MC samples incorporate the CTEQ6.6~\cite{Pumplin:2002vw,Nadolsky:2008zw}
parton distribution functions,
with \PYTHIA used to describe parton showering and hadronization.
The MC distributions account for pileup interactions,
as observed in data.
In addition,
we correct the simulation so that the {\cPqb}-tagging and
misidentification efficiencies match those
determined from control samples in the data.
The {\cPqb}-tagging efficiency correction factor depends
slightly on jet \pt and has a typical value of~0.95~\cite{Chatrchyan:2012jua}.
A further correction,
applied to the signal samples,
accounts for mismodeling of initial-state radiation (ISR) in \MADGRAPH.
The correction is derived by comparing the \pt spectra of reconstructed
{\Z} bosons, \ttbar pairs,
and {\PW}{\Z} pairs between data and simulation.
At high values of transverse momentum of these systems,
where the \pt is balanced by radiated jets,
the \MADGRAPH simulation is found to overestimate the observed event rate.
The corresponding correction is negligible except for small values 
of the gluino-LSP mass difference
where it can be as large as~20\% for both the \tonebbbb and \tonetttt samples.

\section{ Likelihood function and background evaluation methods }
\label{sec:likelihood}

In this section, we present the definition of the likelihood function
and describe the background evaluation methods.
We use the following notation:
\begin{itemize}
  \item  ZL: the zero-lepton event sample;
  \item  SL: the single-lepton event sample;
  \item  LDP: the low-\dphin event sample;
  \item  {\Z}ee and {\Z}$\mu\mu$: the \zee\ and \zmumu\ event samples;
  \item  ttWj:
    the top-quark and \wpjets background component,
    where ``top-quark'' includes both \ttbar and single-top-quark events;
  \item  QCD: the QCD multijet background component;
  \item  $\Z\nu\nu$: the {\Z}+jets (where {\znunu}) background component;
  \item  SUSY: the signal component;
  \item  $\mu^\mathrm{C}_{\mathrm{S};\, i,j,k}$:
    the estimated number of events in bin $i,j,k$
    of event sample S for component C
    without accounting for trigger efficiency,
    where $i$, $j$, and $k$ denote the bin in \met, \HT, and \nbjet, respectively,
    and C denotes ttWj, QCD, or one of the other signal or background terms;
  \item  $n_{\mathrm{S};\, i,j,k}$:
    the estimated number of events in bin $i,j,k$ of event sample S
    from all components after accounting for trigger efficiency;
  \item  $\epsilon^\text{trig}_{\mathrm{S};\, i,j,k}$:
    the trigger efficiency in bin $i,j,k$ for event sample S;
  \item  $N_{\mathrm{S};\, i,j,k}$:
    the observed number of events in bin $i,j,k$ for event sample S.
\end{itemize}

\subsection{Top-quark and \texorpdfstring{\wpjets}{W+jets} background}
\label{sec:ttbarwjets}

The SL sample is used to describe the shape of the top-quark and \wpjets
background in the three analysis dimensions of \met, \HT, and \nbjet.
The SL sample thus provides a three-dimensional (3D)
binned probability density function (PDF) determined directly from data.
The top-quark and \wpjets background in each bin of the ZL sample
is determined from this measured 3D shape,
simulation-derived bin-by-bin corrections $S^{\cPqt\cPqt\PW \Pj}_{i,j,k}$,
and an overall normalization
term $R^{\cPqt\cPqt\PW \Pj}_{\Z \mathrm{L}/\mathrm{SL}}$ that is a free parameter in the fit,
as described below.

With respect to SM processes,
the SL sample is assumed to be populated by top-quark and \wpjets events only.
Contributions from QCD multijet and {\Z}+jets events are small
(around 1\% on average)
as seen from Fig.~\ref{fig:data-mc-compare},
and are accounted for with a systematic uncertainty.
The contribution from \tonebbbb events is negligible
because isolated leptons are rare in the \tonebbbb scenario.
In contrast,
with four top quarks in the final state,
\tonetttt events often contain an isolated high-\pt lepton,
resulting in events that populate the SL sample.
Therefore, we presume
\begin{equation}
  n_{\mathrm{SL};\, i,j,k}  \
   = \ \epsilon^\text{trig}_{\mathrm{SL};\, i,j,k} \cdot (\mu^{\cPqt\cPqt\PW \Pj}_{\mathrm{SL};\, i,j,k}
      + S^\text{SUSY}_{\mathrm{SL};\, i,j,k} \cdot \mu^\text{SUSY}_{\mathrm{SL};\, i,j,k}),
  \label{eq-sl}
\end{equation}
where $S^\text{SUSY}_{\mathrm{SL};\, i,j,k}$ is a nuisance parameter.
For the \tonebbbb scenario,
$\mu^\text{SUSY}_{\mathrm{SL};\, i,j,k}=0$.

\begin{figure*}[tb]
\centering
\includegraphics[width=0.95\linewidth]{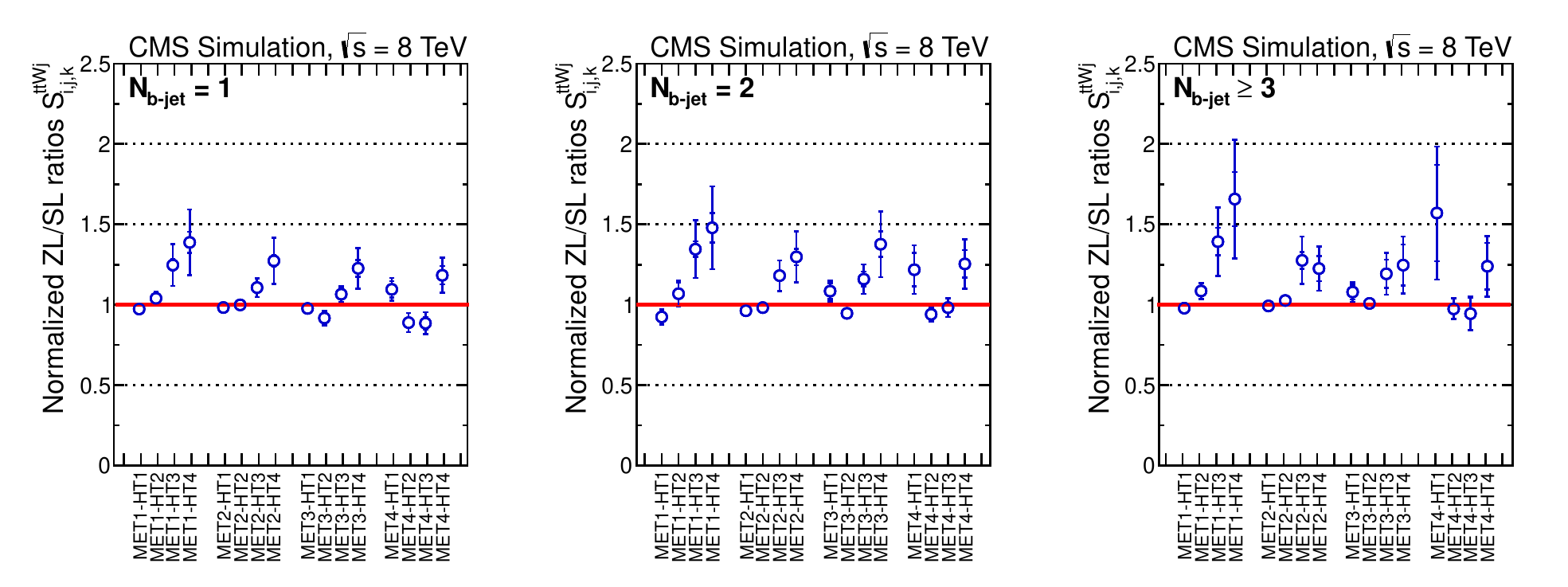}
\caption{
[left] Ratio of the number of events in the zero-lepton (ZL) sample to that
in the single-lepton (SL) sample for simulated top-quark and \wpjets events
in the 16 {\HT}-{\met} bins with $\nbjet=1$,
divided by the average ratio value over the 64 bins with
$\nbjet=1$, 2, and $\geq3$.
The leftmost group of four consecutive points corresponds to
\met bin~1 (MET1) of the table in Fig.~\ref{fig:binning-cartoon},
the next-leftmost group to \met bin~2 (MET2), etc.
The four points within each group correspond to the four \HT bins
in the table,
increasing in \HT value from left to right (HT1 to HT4).
The inner (outer) error bars show the statistical
(combined statistical and systematic) uncertainties.
[center and right]
The corresponding results for $\nbjet=2$ and $\nbjet\geq3$.
}
\label{fig:likelihood-ttwj-closure}
\end{figure*}

We calculate
the ratio of the number of top-quark and \wpjets events in the ZL sample
to the corresponding number in the SL sample,
as predicted by simulation,
after normalization to the same integrated luminosity.
We consider the simulated ZL-to-SL ratios in three groups of 16 bins,
one group corresponding to $\nbjet=1$,
one to $\nbjet=2$,
and one to $\nbjet\geq3$ (see Fig.~\ref{fig:binning-cartoon}).
The 48 ratio values are each normalized by dividing by 
the average ratio value over the 48 bins.
The resulting normalized ZL-to-SL ratios are shown in the left
plot of Fig.~\ref{fig:likelihood-ttwj-closure} for $\nbjet=1$,
in the center plot for $\nbjet=2$,
and in the right plot for $\nbjet\geq3$.
Were the 3D shape of top-quark and \wpjets distributions the same
in the simulated ZL and SL samples,
all points in Fig.~\ref{fig:likelihood-ttwj-closure} would be consistent with unity.
Deviations from unity on the order of 20--50\% are seen for some points,
indicating a shape difference between the two samples.
The shape difference is strongest in the \HT dimension.
This \HT dependence is due to the lepton isolation requirement,
which is less likely to be satisfied as \HT increases.
Consistent results are found if the {\POWHEG} or
{\MCATNLO}~\cite{Frixione:2002ik} generator,
rather than {\MADGRAPH},
is used to produce the \ttbar MC sample.

Our estimate of the top-quark and \wpjets contribution to bin $i,j,k$
of the ZL sample is thus
\begin{equation}
  \mu^{\cPqt\cPqt\PW \Pj}_{\mathrm{ZL};\, i,j,k} \
    = \ S^{\cPqt\cPqt\PW \Pj}_{i,j,k} \cdot R^{\cPqt\cPqt\PW \Pj}_{\mathrm{ZL}/\mathrm{SL}} \cdot \mu^{\cPqt\cPqt\PW \Pj}_{\mathrm{SL};\, i,j,k},
  \label{eq-zl}
\end{equation}
where $R^{\cPqt\cPqt\PW \Pj}_{\mathrm{ZL}/\mathrm{SL}}$ is the scale factor
common to all bins mentioned above
and the $S^{\cPqt\cPqt\PW \Pj}_{i,j,k}$ factors are the MC-based terms
presented in Fig.~\ref{fig:likelihood-ttwj-closure},
which account for
the 3D shape differences between the ZL and SL samples.
In the likelihood function,
the $S^{\cPqt\cPqt\PW \Pj}_{i,j,k}$ terms are treated as nuisance parameters
whose values are determined in the fit,
each constrained by a lognormal PDF.
The median of the lognormal is the corresponding value
shown in Fig.~\ref{fig:likelihood-ttwj-closure},
while the geometric standard deviation is $\ln(1+\sigma_\text{rel})$,
with $\sigma_\text{rel}$ the relative uncertainty of the
corresponding $S^{\cPqt\cPqt\PW \Pj}_{i,j,k}$ term,
determined from the quadratic sum of its statistical uncertainty and
one half the difference from unity.
In addition,
we vary the \wpjets cross section by 100\%~\cite{Aad:2011kp}.
The difference with respect to the standard result defines an
uncertainty for a lognormal distribution that is applied as
an additional constraint on the $S^{\cPqt\cPqt\PW \Pj}_{i,j,k}$ terms.
An analogous constraint is derived through variation of the
single-top-quark cross section by 30\%~\cite{Chatrchyan:2012yva}.

\subsection{ QCD multijet background }
\label{sec:qcd}

The QCD multijet background in each bin of the ZL sample,
in the 3D space of \met, \HT, and \nbjet,
is determined from the number of events in the corresponding bin
of the LDP sample,
in conjunction with multiplicative scale factors described below.
Before applying these scale factors,
the contributions of top-quark and \wpjets events
are subtracted from the measured LDP results,
as are the contributions of {\Z}+jets events.
The estimate of the top-quark and \wpjets contribution
to the LDP sample is determined from the
data-derived top-quark and \wpjets event yield in the ZL sample,
found in the likelihood fit (Section~\ref{sec:results}) for the corresponding bin,
multiplied by the MC ratio of LDP to ZL events for that bin,
and analogously for the {\Z}+jets contribution to the LDP sample
(these subtractions are performed simultaneously with all other
aspects of the fit).
The uncertainty assigned to this subtraction procedure accounts for
the total uncertainty of the respective ZL event yield,
and for a 10\% uncertainty associated with the simulated ratio,
where the latter term corresponds to the average statistical
uncertainty of the MC ratio values.

\begin{figure*}[tb]
\centering
\includegraphics[width=0.99\linewidth]{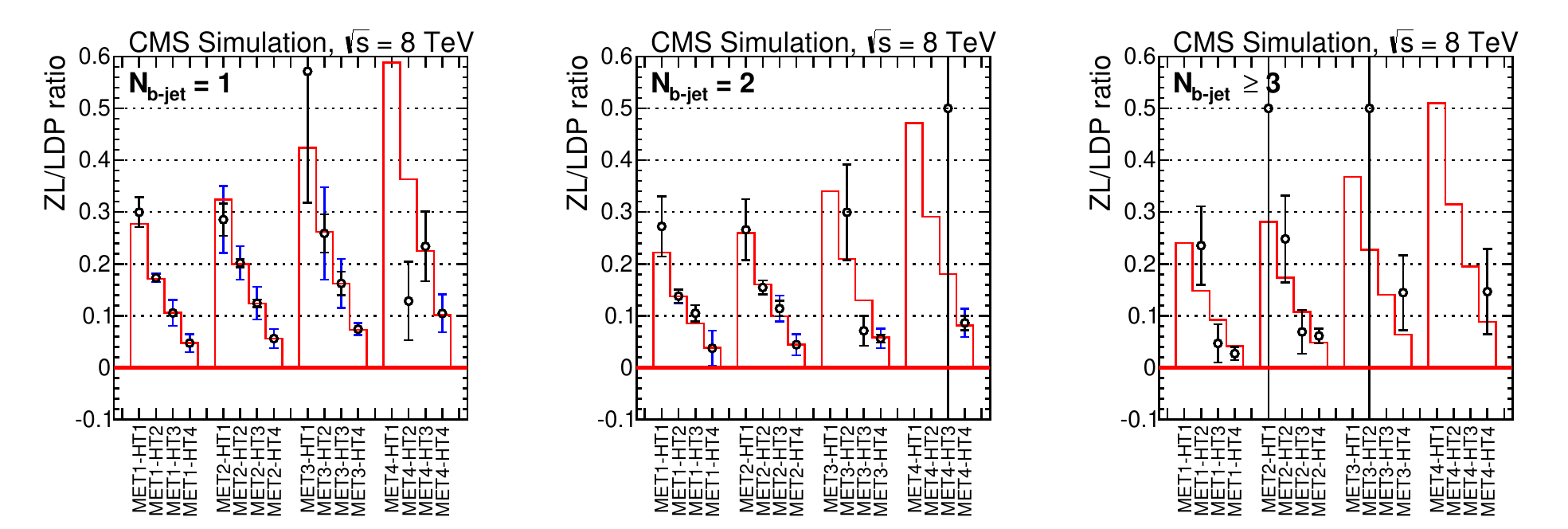}
\includegraphics[width=0.99\linewidth]{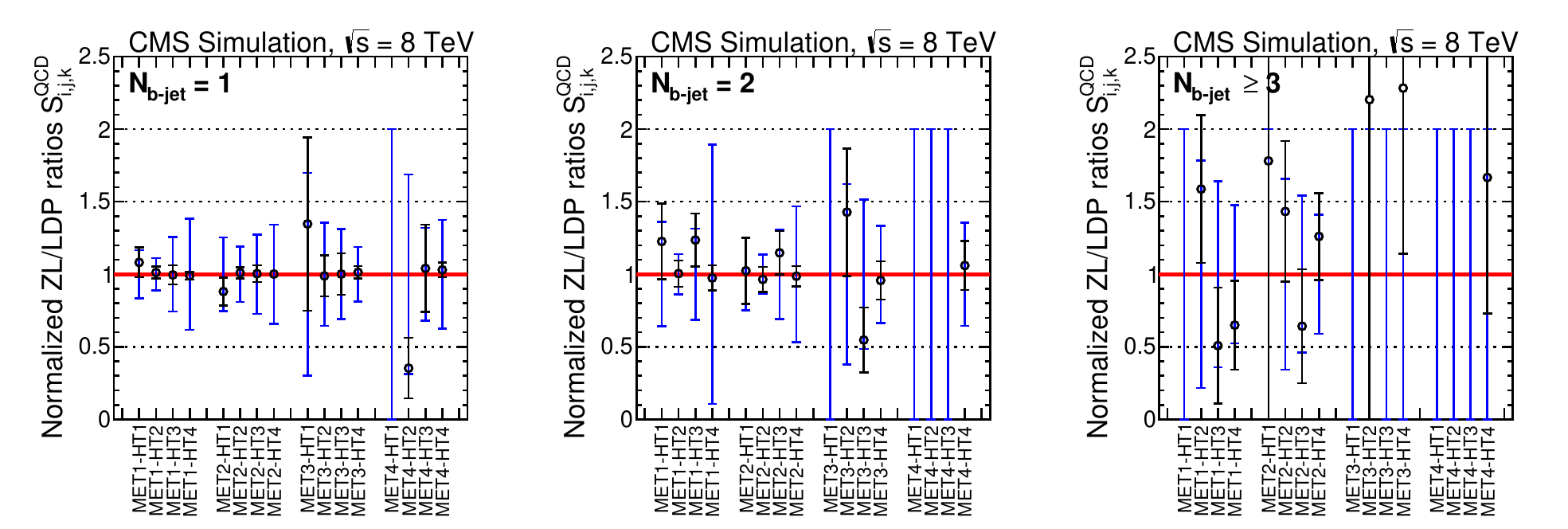}
\caption{
[top row]
Ratio of the number of events in the zero-lepton (ZL) sample to that
in the low-\dphin (LDP) sample for simulated QCD multijet events.
The definitions of the bins are the same as
in Fig.~\ref{fig:likelihood-ttwj-closure}.
Various QCD multijet samples,
with different choices for the hardness scale
($\hat{p}_{\mathrm{T}}$~\cite{bib-pythia})
of the interaction,
are combined.
The points show the averages over those samples.
The inner error bars indicate the statistical uncertainties.
The outer error bars indicate the statistical uncertainties
added in quadrature with the
root-mean-squared values over the different $\hat{p}_{\mathrm{T}}$ samples.
The histogram shows the results of the fitted parameterization described
in the text.
[bottom row]
The corresponding ratio divided by the
parameterization from the top row.
The inner (black) and outer (blue) error bars indicate the statistical
and combined statistical-and-systematic uncertainties,
respectively.
}
\label{fig:likelihood-qcd-closure}
\end{figure*}

The top row of Fig.~\ref{fig:likelihood-qcd-closure} shows the ratio
between the number of QCD multijet events in the ZL sample to the
corresponding number in the LDP sample,
as predicted by simulation,
after normalization to the same integrated luminosity.
The results are shown for the 48 bins of the ZL and LDP samples.
This ratio is seen to depend strongly on \HT.
The dependence on \met and \nbjet is more moderate.
We parameterize the \met, \HT, and \nbjet dependence
assuming that this dependence factorizes,
\ie, we assume that the \HT dependence is independent of \met and \nbjet, etc.
We thus model the QCD multijet background contribution to the ZL sample for a given
\met, \HT, \nbjet bin as:
\begin{equation}
  \mu^\mathrm{QCD}_{\mathrm{ZL};\, i,j,k} = S^\mathrm{QCD}_{i,j,k} \cdot
          \left( K^\mathrm{QCD}_{\mathrm{MET},i}
           \cdot K^\mathrm{QCD}_{\mathrm{HT},j} \cdot K^\mathrm{QCD}_{\mathrm{Nb},k} \right) \cdot
          \mu^\mathrm{QCD}_{\mathrm{LDP};\, i,j,k},
\label{eq-qcd-param}
\end{equation}
where the three $K^\mathrm{QCD}$ terms describe the
\met, \HT, and \nbjet dependence and
the $S^\mathrm{QCD}_{i,j,k}$ factors (defined below) are corrections
to account for potential inadequacies in the parametrization.
Note that some bins in the top row of
Fig.~\ref{fig:likelihood-qcd-closure} do not contain any entries.
These bins generally have large \met and small \HT values,
making them kinematically unlikely
(a large \met value implies a large \HT value),
and thus contain few or no events.

We fit the parameterization of Eq.~(\ref{eq-qcd-param}) to the
ratio values shown in the top row of
Fig.~\ref{fig:likelihood-qcd-closure},
taking $S^\mathrm{QCD}_{i,j,k}\equiv 1$ at this stage,
to determine simulation-derived values for the $K^\mathrm{QCD}$ factors
(for the final results,
most $K^\mathrm{QCD}$ factors are determined in the likelihood fit,
as explained below).
The results of this fit are shown by the histograms in
the top row of Fig.~\ref{fig:likelihood-qcd-closure}.
The simulated QCD ZL-to-LDP ratios divided by the fitted parameterization
are shown in the bottom row of Fig.~\ref{fig:likelihood-qcd-closure}.
The points in the bottom row are consistent with unity,
indicating that the empirical parameterization of Eq.~(\ref{eq-qcd-param}) is sufficient.
Therefore, in the likelihood fit,
no corrections to the parametrization are applied.
The $S^\mathrm{QCD}_{i,j,k}$ factors are treated as nuisance parameters
constrained by lognormal PDFs with a median set to unity.
Geometric standard deviations for the lognormal distributions
are set equal to the outer error bars in the bottom row of
Fig.~\ref{fig:likelihood-qcd-closure},
given by the quadratic
sum of the deviation of the ratios in the
bottom row of Fig.~\ref{fig:likelihood-qcd-closure} from unity,
the statistical uncertainties of these ratios,
and the root-mean-squared values found using the different QCD multijet samples
described in the Fig.~\ref{fig:likelihood-qcd-closure} caption.
For bins in the top row of Fig.~\ref{fig:likelihood-qcd-closure}
without any MC entries,
we assign 100\% uncertainties,
which are indicated in the bottom row of the figure.

In the likelihood analysis,
most of the $K^\mathrm{QCD}$ factors are free parameters in the fit:
there is enough shape information that they
can be determined directly from the data.
However,
we find from studies with simulation
that the fit is unable to determine
$K^\mathrm{QCD}_{\mathrm{MET},3}$, $K^\mathrm{QCD}_{\mathrm{MET},4}$, or $K^\mathrm{QCD}_{\mathrm{Nb},3}$.
Instead,
lognormal constraints are applied for these three parameters.
The median values are set to the corresponding results from simulation
and the geometric standard deviations to half the differences
$K^\mathrm{QCD}_{\mathrm{MET},3}-K^\mathrm{QCD}_{\mathrm{MET},2}$,
$K^\mathrm{QCD}_{\mathrm{MET},4}-K^\mathrm{QCD}_{\mathrm{MET},2}$,
and $K^\mathrm{QCD}_{\mathrm{Nb},3}-K^\mathrm{QCD}_{\mathrm{Nb},1}$, respectively.
The results of the fit are found to be insensitive
to the choice of the geometric standard deviation values.

\subsection{ The \texorpdfstring{{\Z}+jets}{Z+jets} background }
\label{sec:like-znunu}

The {\Z}+jets background (where \znunu) is evaluated by
reconstructing \zll\ events ($\ell =\,${\Pe} and {\Pgm}).
The $\ell^+$ and $\ell^-$ leptons are then removed so that the events
emulate {\Z}+jets events with {\znunu.}
The \zee and \zmumu samples are divided into 16 bins in
the two-dimensional space of \met and \HT,
as indicated in Fig.~\ref{fig:binning-cartoon}.

Fits to the dilepton invariant mass spectra
are performed to determine the \zll\ yields.
The yields are corrected to account for background,
acceptance, and detection efficiency.
The acceptance,
determined from simulation,
accounts for the larger fiducial volume
for the detection of \znunu events compared to \zll\ events.
The efficiency is
$\epsilon = \epsilon_{\mathrm{trig}}
  \cdot \epsilon_{\mathrm{\ell\, reco}}^2
  \cdot \epsilon_{\mathrm{\ell\, sel}}^2$,
where the trigger $\epsilon_{\mathrm{trig}}$,
lepton reconstruction $\epsilon_{\mathrm{\ell\, reco}}$,
and lepton selection $\epsilon_{\mathrm{\ell\, sel}}$
factors are determined from data.

The \zll\ yields are small in some of the signal regions.
To increase these yields,
we select events with the requirements of Section~\ref{sec:basiccuts}
except with a significantly looser \cPqb-tagging definition.
The yield in each bin of this sample is multiplied by an
extrapolation factor
given by the ratio of the sum of the \zll\ yields
over all \HT and \met bins for events that satisfy the nominal
\cPqb-tagging requirements
to those that satisfy the loose requirements.

To establish whether the extrapolation factors themselves exhibit a
dependence on \HT\ or \met,
we construct a control sample identical to
the LDP sample except with the loosened {\cPqb}-tagging definition.
This sample is dominated by QCD multijet production,
and is found to have a distribution for the output variable of the
{\cPqb}-tagging algorithm similar to that of the \zll\ events.
From this control sample,
we find that the $\nbjet=1$ extrapolation factors exhibit a
variation with \met\ up to 25\%;
we apply this variation as a correction to those factors.
For $\nbjet=2$ and $\nbjet\geq 3$,
we find no variation within
the uncertainties and do not apply a correction.

The {\Z}+jets background in the $i=\met$, $j=\HT$ bin of the ZL sample
with $\nbjet=1$ is related to the corresponding bin
in the \zee and \zmumu control samples through
\ifthenelse{\boolean{cms@external}}
{
\begin{align}
\begin{split}
\label{eqn:zee}
  \mu^{\Z\Pe\Pe}_{{\Z}{\Pe\Pe};i,j}  = &
     \left( \mu^{\Z\nu\nu}_{\mathrm{ZL};\, i,j,1} \cdot S_{\Pe\Pe} \cdot A_{{\Pe\Pe};\, i} \cdot \epsilon_{\Pe\Pe}\right) /\\
     &\left( \mathcal{F}_{\Z\nu\nu;\, 1} \cdot R_\mathrm{B}\right),
\end{split}\\
\begin{split}
\label{eqn:zmm}
  \mu^{\Z\mu\mu}_{\Z\mu\mu;i,j}  = &
     \left( \mu^{\Z\nu\nu}_{\mathrm{ZL};\, i,j,1} \cdot S_{\mu\mu} \cdot A_{\mu\mu;\, i} \cdot \epsilon_{\mu\mu}\right) /\\
                &\left( \mathcal{F}_{\Z\nu\nu;\, 1} \cdot R_\mathrm{B}\right),
\end{split}
\end{align}
}
{
\begin{align}
\label{eqn:zee}
  \mu^{\Z\Pe\Pe}_{{\Z}{\Pe\Pe};i,j}   = &
     \left( \mu^{\Z\nu\nu}_{\mathrm{ZL};\, i,j,1} \cdot S_{\Pe\Pe} \cdot A_{{\Pe\Pe};\, i} \cdot \epsilon_{\Pe\Pe}\right) /
     \left( \mathcal{F}_{\Z\nu\nu;\, 1} \cdot R_\mathrm{B}\right), \\
\label{eqn:zmm}
  \mu^{\Z\mu\mu}_{\Z\mu\mu;i,j}  = &
     \left( \mu^{\Z\nu\nu}_{\mathrm{ZL};\, i,j,1} \cdot S_{\mu\mu} \cdot A_{\mu\mu;\, i} \cdot \epsilon_{\mu\mu}\right) /
                \left( \mathcal{F}_{\Z\nu\nu;\, 1} \cdot R_\mathrm{B}\right),
\end{align}
}
where $A_{\ell\ell;\, i}$ and $\epsilon_{\ell\ell}$
are the acceptances and efficiencies for the \zll samples, respectively,
$S_{\ell\ell}$ is a scale factor to account for systematic uncertainties,
$R_\mathrm{B}=5.95\pm 0.02$ is the ratio of the \znunu
and \zll branching fractions~\cite{Beringer:1900zz},
and $\mathcal{F}_{Z\nu\nu;\, 1}$ is the extrapolation factor that relates
the $\nbjet = 1$ selection efficiency to the efficiency
of the loose \cPqb-tagging requirement.
The estimates of the {\Z}+jets background for
$\nbjet =2$ and $\nbjet \ge 3$
are given by the \mbox{$\nbjet=1$} result through the ratio of
\cPqb-tagging extrapolation factors:
\begin{equation}
\label{eqn:znn23}
  \mu^{{\Z}\nu\nu}_{\mathrm{ZL};\, i,j,k}  =  \mu^{{\Z}\nu\nu}_{\mathrm{ZL};\, i,j,1} \cdot
\left( \mathcal{F}_{{\Z}\nu\nu;\, k} / \mathcal{F}_{{\Z}\nu\nu;\, 1}  \right),
\end{equation}
where $k$ is the \nbjet bin index.

Systematic uncertainties are evaluated
for the \zll\ purity, acceptance, and detection efficiency
by considering their dependence on \met and~\HT,
and by varying the selection conditions.
An additional uncertainty,
based on a consistency test performed with simulation,
accounts for the level of agreement between the predicted and
correct \znunu event rates.
Finally,
systematic uncertainties are evaluated for the extrapolation factors
by varying the loosened {\cPqb}-tagging definition
and by assigning an uncertainty to account for
the observed or potential variation with \met and~\HT
(for the $\nbjet=2$ and $\nbjet\geq 3$ factors,
the latter uncertainty is based on the level of statistical fluctuation).
The total systematic uncertainty of the \znunu background estimate
is 30\% for $\nbjet=1$,
35\% for $\nbjet=2$,
and 60\% for $\nbjet\geq3$.

\subsection{ Other backgrounds }

Backgrounds from diboson and Drell-Yan
processes are accounted for using simulation,
with an uncertainty of~100\%.
Their total fractional contribution to the overall background
is 1\% or less in all search regions.

\subsection{ Systematic uncertainties }
\label{subsec:likelihood-systematics}

Systematic uncertainties associated with the signal efficiency arise from various sources.
A systematic uncertainty associated with the jet energy scale is
evaluated by varying this scale by its {\pt}- and $\eta$-dependent uncertainties.
The size of this uncertainty depends on the event kinematics,
i.e., the \met bin, the \HT bin,
and the assumed values of the gluino and LSP masses:
typical values are in the range of 5--10\%.
A systematic uncertainty of 1\% is associated with unclustered energy.
This uncertainty is evaluated by varying the transverse energy in
an event not clustered into a physics object by~10\%.
A systematic uncertainty of 3\% is associated with anomalous \met values,
caused by events that
are misreconstructed or that contain beam-related background.
This uncertainty is defined by 100\% of the change in efficiency when
software filters are applied to reject these events.
The uncertainty of the luminosity determination is 4.4\%~\cite{CMS-PAS-LUM-12-001}.
The systematic uncertainties associated with corrections to
the jet energy resolution,
the pileup modeling mentioned in Section~\ref{sec:basiccuts},
the trigger efficiency,
the \cPqb-tagging efficiency scale factor,
and the ISR modeling
are evaluated by varying the respective quantities by their uncertainties,
while systematic uncertainties associated with the parton distribution functions
are evaluated~\cite{Pumplin:2002vw,Martin:2009iq,Ball:2011mu}
following the recommendations of Ref.~\cite{PDF4LHC}.
The jet energy resolution and pileup modeling uncertainties are
2\% and 3\%, respectively.
The uncertainty of the trigger efficiency is generally below~2\%.
Uncertainties associated with the parton distribution functions
and \cPqb-tagging efficiency
are typically below 10\% and 15\%, respectively.
The uncertainties of the \tonebbbb (\tonetttt) ISR modeling
corrections are typically 5\% (3\%),
but can be as large as 20\% (20\%) near the $\mgluino = \mlsp$ diagonal.
The uncertainties associated with the jet energy scale,
\cPqb-tagging efficiency,
ISR modeling,
and parton distribution functions
vary significantly
with the event kinematics and are evaluated point-by-point in the scans
over gluino and LSP masses discussed in Section~\ref{sec:results}.

Systematic uncertainties for the SM background estimates
are described in the previous sections.
Note that,
for our analysis,
systematic uncertainties are generally much smaller than
statistical uncertainties,
where the latter terms primarily arise as a consequence of
the limited numbers of events in the data control samples.

\subsection{ The global likelihood function }
\label{subsec:likelihood-global}

The likelihood function is the product of Poisson PDFs,
one for each bin,
and the constraint PDFs for the nuisance parameters.
For each bin,
the Poisson PDF gives the probability to observe
$N$ events, given a mean $n$,
where $n$ depends on the parameters of the likelihood model such as those given
in Eqs.~(\ref{eq-sl})--(\ref{eqn:znn23}).
The region with $\met>350\GeV$ and $400<\HT<500\GeV$,
representing the bin with highest \met and lowest \HT in our analysis
(the HT1-MET4 bin of Fig.~\ref{fig:binning-cartoon}),
is at an extreme limit of phase space and is very sparsely populated,
making it difficult to validate the background evaluation procedures.
Furthermore,
very few signal events are expected in this region.
We therefore exclude the HT1-MET4 bin from the likelihood analysis,
corresponding to 11 of the 176 bins.
Thus,
the effective number of bins in the analysis is~165.

For both signal and background terms,
external input parameters are allowed to vary and are
constrained by a PDF in the likelihood.
Parameters with values between zero and one,
such as efficiencies,
are constrained by beta-distribution PDFs
(see Section 35 of Ref.~\cite{Beringer:1900zz}).
All others are constrained by lognormal PDFs.
Correlations between the different kinematic regions,
including the \nbjet bins,
are taken into account.
The test statistic is
$q_\mu =  - 2 \ln \left( {\cal L}_\mu/{\cal L}_\text{max} \right)$,
where ${\cal L}_\text{max}$ is the maximum likelihood
determined by allowing all parameters including the SUSY signal strength $\mu$ to vary,
and ${\cal L}_\mu$ is the maximum likelihood for a fixed signal strength.

\section{Results}
\label{sec:results}

\begin{figure*}[tbhp]
\centering
\includegraphics[width=0.90\linewidth]{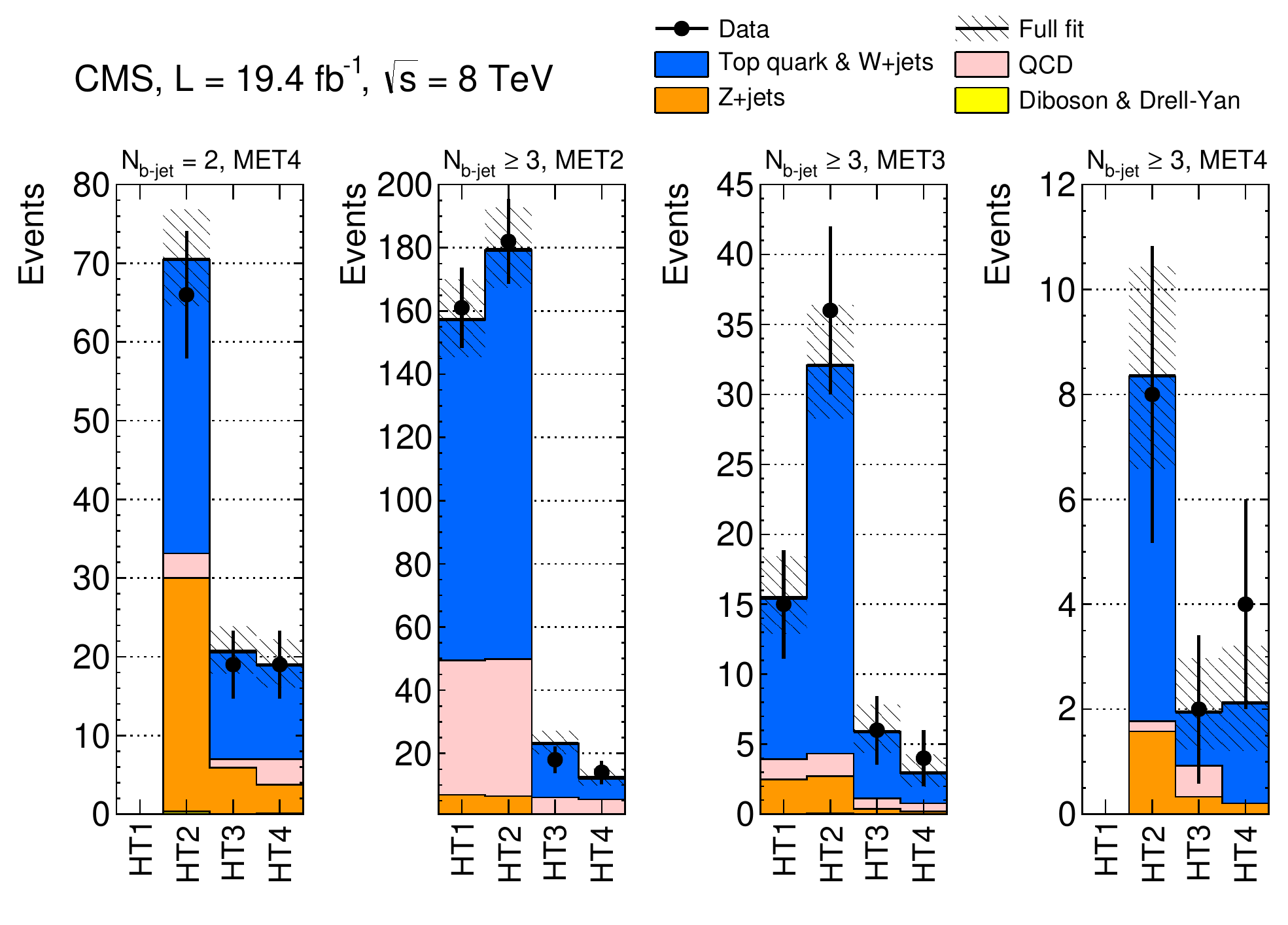}
\caption{
Observed numbers of events (points with error bars)
for the 14 bins with highest signal sensitivity in the analysis,
in comparison with the standard model background predictions
(with total uncertainties shown by the hatched bands)
found in the fit
with SUSY signal strength fixed to zero.
The labels HT1, HT2, MET2, etc., refer to the bins of
\HT and \met indicated in Fig.~\ref{fig:binning-cartoon}.
}
\label{fig:fitresults-hsbins}
\end{figure*}

SUSY events in the \tonebbbb and \tonetttt scenarios
often contain significant \met and multiple \cPqb\ jets,
as discussed in the Introduction.
Tables~\ref{tab:pred-and-observed-nb2}
and~\ref{tab:pred-and-observed-nb3}
and Fig.~\ref{fig:fitresults-hsbins}
present the results of the fit for the
14 bins of the analysis that we find to be most sensitive to these two scenarios:
the three bins with $\HT>500\GeV$,
$\met>350\GeV$,
and $\nbjet=2$,
for which the results are shown in
Table~\ref{tab:pred-and-observed-nb2},
and the 11 bins with $\met>150\GeV$ and $\nbjet\geq3$,
for which the results are shown in
Table~\ref{tab:pred-and-observed-nb3}.
For these results,
the SUSY signal strength is set to zero
so that we can test the compatibility of the data with the SM hypothesis.
For the scan results over gluino and LSP masses presented below,
the SUSY signal strength is allowed to vary.

The top row of Table~\ref{tab:pred-and-observed-nb2}
and top section of Table~\ref{tab:pred-and-observed-nb3}
show the numbers of events observed in data.
The second row and section show the SM background estimates
obtained from the fit,
which are seen to be in agreement
with the data to within the uncertainties.
The third row and section present the SM
predictions from the simulation.
The simulated results are for guidance only and are not
used in the analysis.

\begin{table*}[tb]
\begin{center}
\topcaption{
Observed numbers of events,
SM background estimates from the fit,
and SM expectations from Monte Carlo simulation,
for the signal (ZL) regions with $\met>350\GeV$ and $\nbjet=2$.
The labels HT2, HT3, and HT4 refer to the bins of
\HT indicated in Fig.~\ref{fig:binning-cartoon},
while HT2-4 is the sum over the three bins.
The fourth row presents the SM background estimates from the sideband fit
described in the text.
The uncertainties listed for the fit results include
the statistical and systematic components,
while those shown for the simulation are statistical only.
For the fits,
the SUSY signal strength is fixed to zero.
The last row shows the expected numbers of events from a
SUSY test scenario described in the text.
}
\begin{tabular}{lcccc}
\hline
  $\nbjet=2$, MET4 &  HT2 & HT3 & HT4 & HT2-4 \vsmtvs \\
\hline
  Observed number of events          &  66  & 19  & 19  &  104  \vsmtvs \\
  SM background estimates from fit & $70.5\ ^{+6.3}_{-5.9}$    & $20.7\ ^{+3.2}_{-2.8}$    & $19.0\ ^{+3.2}_{-2.8}$    & $110\pm 8$  \vsmtvs \\
  SM background predictions from simulation  &    $81.6 \pm 1.9$  &       $28.7 \pm 1.3$  &    $23.3 \pm 0.8$  &    $134 \pm 2$  \vsmtvs \\
  SM background estimates from sideband fit & $76.4\ ^{+10.2}_{-9.1}$    & $22.3\ ^{+4.5}_{-3.9}$    & $19.0\ ^{+4.5}_{-3.7}$    & $118\ ^{+13}_{-12}$  \vsmtvs \\
Number of signal events, SUSY test scenario & 0.5  &  1.5 &  11.6 &  13.6 \vsmtvs \\
\hline
\end{tabular}
\label{tab:pred-and-observed-nb2}
\end{center}
\end{table*}

\begin{table*}[tpbh]
\begin{center}
\topcaption{
Observed numbers of events,
SM background estimates from the fit,
and SM expectations from Monte Carlo simulation,
for the signal (ZL) regions with $\met>150\GeV$ and $\nbjet\geq 3$.
The labels HT1, HT2, MET2, etc., refer to the bins of
\HT and \met indicated in Fig.~\ref{fig:binning-cartoon},
while
HT1-4 (MET2-4) is the sum over the four \HT (three \met) bins.
The HT1-MET4 bin is excluded from the analysis,
as explained in the text.
The fourth section presents the SM background estimates from the sideband fit
described in the text.
The uncertainties listed for the fit results include
the statistical and systematic components,
while those shown for the simulation are statistical only.
For the fits,
the SUSY signal strength is fixed to zero.
The last section shows the expected numbers of events from a
SUSY test scenario described in the text.
}
\begin{tabular}{cccccc}
\hline
\multicolumn{6}{c}{ Observed number of events } \\
\hline
 $\nbjet\ge3$  & HT1 & HT2 & HT3 & HT4 & HT1-4 \\
\hline
  MET2  &  161  &  182  &  18  &  14  &  375  \vsmtvs \\
  MET3  &  15  &  36  &  6  &  4  &  61  \vsmtvs \\
  MET4  &  ---  &  8  &  2  &  4  &  14  \vsmtvs \\
    \hline
  MET2-4   &  176  &  226  &  26  &  22  &  450  \vsmtvs \\
    \hline
    \hline
    \multicolumn{6}{c}{ SM background estimates from fit } \\
    \hline
 $\nbjet\ge3$ & HT1 & HT2 & HT3 & HT4 & HT1-4 \\
    \hline
  MET2  &  $157\ ^{+13}_{-12}$    &  $179\ ^{+13}_{-12}$    &  $23.2\ ^{+3.8}_{-3.4}$    &  $12.3\ ^{+2.7}_{-2.3}$    &  $372\ ^{+19}_{-18}$   \vsmtvs \\
  MET3  &  $15.5\ ^{+3.0}_{-2.6}$    &  $32.1\ ^{+4.3}_{-3.8}$    &  $5.9\ ^{+1.9}_{-1.5}$    &  $2.9\ ^{+1.3}_{-1.0}$    &  $56.5\ ^{+5.7}_{-5.4}$   \vsmtvs \\
  MET4  &  ---    &  $8.4\ ^{+2.1}_{-1.8}$    &  $2.0\ ^{+1.0}_{-0.7}$    &  $2.1\ ^{+1.1}_{-0.9}$    &  $12.4\ ^{+2.5}_{-2.2}$   \vsmtvs \\
    \hline
  MET2-4&  $173\ ^{+13}_{-12}$    &  $220\ ^{+14}_{-13}$    &  $31.0\ ^{+4.3}_{-3.8}$    &  $17.3\ ^{+3.1}_{-2.8}$    &  $441\ ^{+20}_{-19}$   \vsmtvs \\
    \hline
    \hline
    \multicolumn{6}{c}{ SM background predictions from simulation } \\
    \hline
 $\nbjet\ge3$ & HT1 & HT2 & HT3 & HT4 & HT1-4 \\
    \hline
  MET2  &  $127 \pm 8$    &  $180 \pm 12$  &  $27 \pm 2$  &  $13 \pm 1$  &  $347 \pm 14$ \vsmtvs \\
  MET3  &  $14.7 \pm 0.7$    &  $30.9 \pm 0.7$  &  $7.5 \pm 0.4$  &  $3.9 \pm 0.2$  &  $56.9 \pm 2.6$ \vsmtvs \\
  MET4  &  ---    &  $6.1 \pm 0.2$  &  $2.6 \pm 0.2$  &  $2.6 \pm 0.2$  &  $11.3 \pm 0.3$ \vsmtvs \\
    \hline
  MET2-4  &  $141 \pm 8$    &  $217 \pm 12$  &  $37 \pm 2$  &  $20 \pm 1$  &  $415 \pm 15$ \vsmtvs \\
    \hline
    \hline
    \multicolumn{6}{c}{ SM background estimates from sideband fit } \\
    \hline
 $\nbjet\ge3$ & HT1 & HT2 & HT3 & HT4 & HT1-4 \\
    \hline
  MET2  &  $119\ ^{+32}_{-19}$    &  $158\ ^{+36}_{-24}$    &  $28.2\ ^{+6.9}_{-5.7}$    &  $10.2\ ^{+3.5}_{-2.7}$    &  $316\ ^{+49}_{-37}$   \vsmtvs \\
  MET3  &  $15.2\ ^{+4.3}_{-3.5}$    &  $27.7\ ^{+5.8}_{-4.9}$    &  $5.6\ ^{+2.6}_{-1.9}$    &  $2.0\ ^{+1.5}_{-0.9}$    &  $50.5\ ^{+8.2}_{-7.3}$   \vsmtvs \\
  MET4  &  ---    &  $8.3\ ^{+2.9}_{-2.2}$    &  $1.9\ ^{+1.3}_{-0.8}$    &  $0.4\ ^{+0.6}_{-0.2}$    &  $10.5\ ^{+3.2}_{-2.5}$   \vsmtvs \\
    \hline
  MET2-4&  $134\ ^{+32}_{-20}$    &  $194\ ^{+36}_{-26}$    &  $35.7\ ^{+7.5}_{-6.3}$    &  $12.6\ ^{+3.8}_{-3.0}$    &  $377\ ^{+51}_{-42}$   \vsmtvs \\
    \hline
    \hline
    \multicolumn{6}{c}{ Number of signal events, SUSY test scenario } \\
    \hline
 $\nbjet\ge3$ & HT1 & HT2 & HT3 & HT4 & HT1-4 \\
    \hline
MET2 & 0.0 & 0.1 & 0.2 & 1.0 & 1.4 \vsmtvs \\
MET3 & 0.0 & 0.2 & 0.4 & 2.0 & 2.6 \vsmtvs\\
MET4 &  ---   & 0.4 & 1.4 & 10.8 & 12.6 \vsmtvs \\
MET2-4 & 0.0 & 0.7 & 2.0 & 13.8 & 16.6 \vsmtvs \\
    \hline
\end{tabular}
\label{tab:pred-and-observed-nb3}
\end{center}
\end{table*}

It is also interesting to perform the likelihood fit
with the Poisson PDF terms for the 14 ``most sensitive'' bins removed,
in order to ascertain the data-derived SM background estimates
when the data in these bins do not affect the result.
We call such a fit the ``sideband'' fit,
which is therefore based on 151 bins.
The sideband fit results for the numbers of SM background events
in the 14 bins are presented in the
fourth row of Table~\ref{tab:pred-and-observed-nb2}
and section of Table~\ref{tab:pred-and-observed-nb3}.
For the sideband fit,
the deviations with respect to the data are seen to be somewhat larger
than for the standard fit.
The largest deviation between observation and SM expectation occurs for
the bin with $\nbjet\geq3$,
$\HT>1000\GeV$, and $\met>350\GeV$
(the HT4-MET4 bin of Table~\ref{tab:pred-and-observed-nb3}),
where 4~events are observed whereas only $0.4^{+0.6}_{-0.2}$ events are expected
(note that these uncertainties are not Gaussian).
From studies with ensembles of simulated experiments,
considering only this bin,
we estimate the probability for a fluctuation in the background in this bin
to match or exceed 4~events to be 9\%
and do not consider this excess further.

For purposes of illustration,
the last row of Table~\ref{tab:pred-and-observed-nb2}
and section of Table~\ref{tab:pred-and-observed-nb3}
show the expected numbers of signal events for a \tonebbbb
"test scenario" near the limit of our sensitivity,
with $\mgluino = 1225\GeV$ and  $\mlsp=150\GeV$.

\begin{figure*}[tbhp]
\centering
\includegraphics[width=\cmsFigWidthTwo]{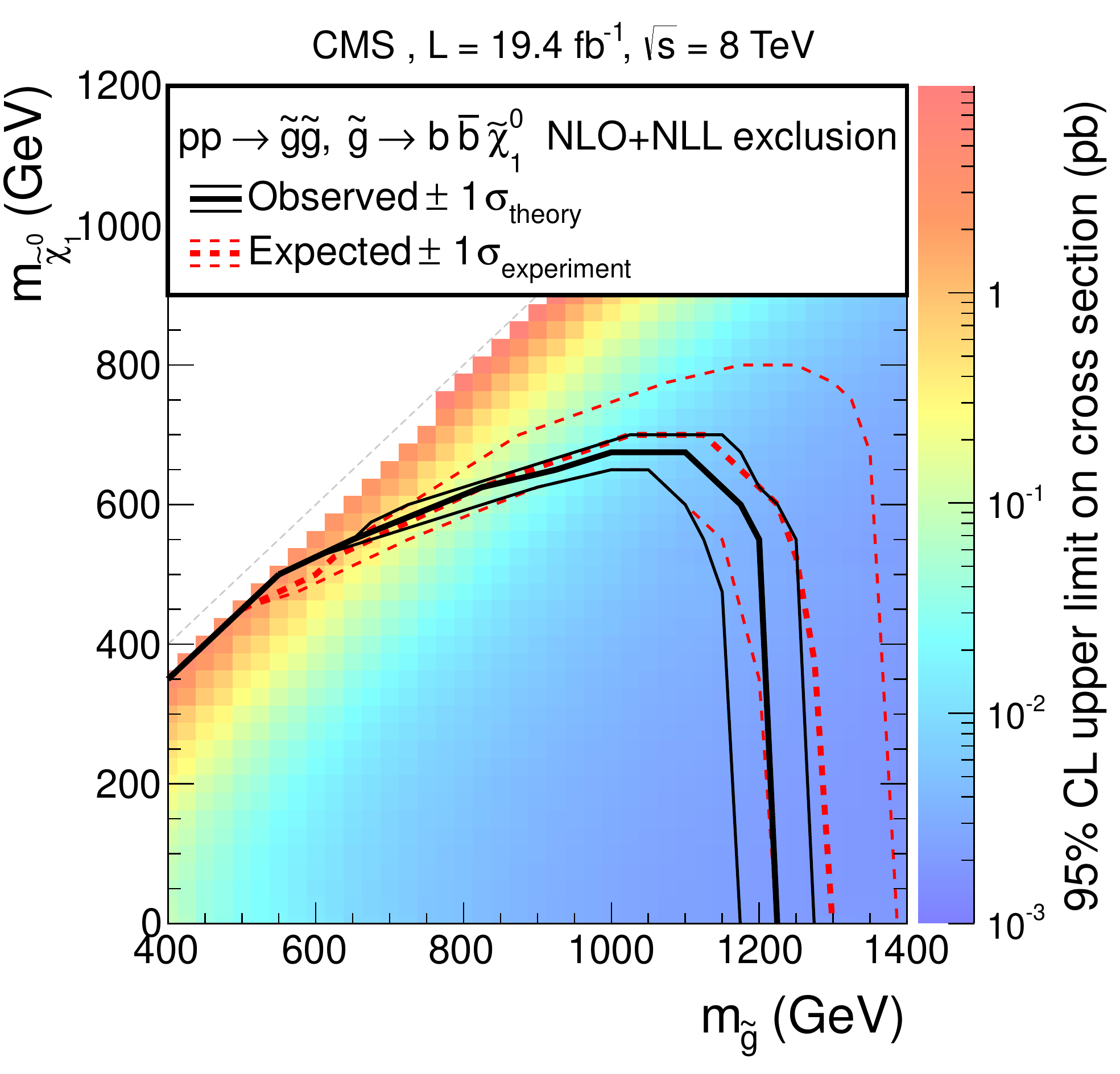}
\includegraphics[width=\cmsFigWidthTwo]{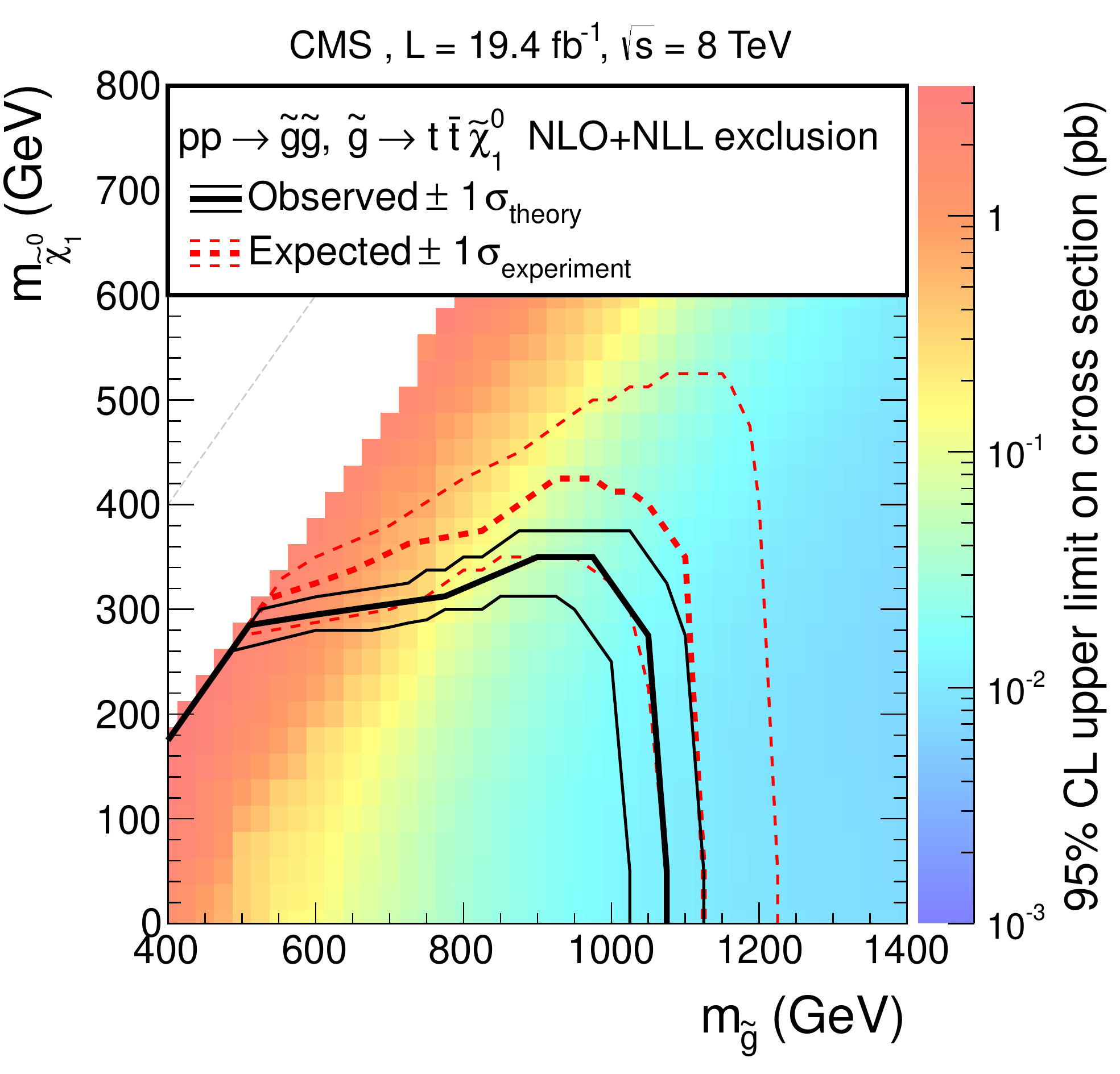}
\caption{
The 95\% CL upper limits on the [left]~\tonebbbb
and [right]~\tonetttt new-physics scenario cross sections (pb)
derived using the CL$_\mathrm{s}$ method.
The solid (black) contours show the observed exclusions
assuming the NLO+NLL cross
sections~\cite{bib-nlo-nll-01,bib-nlo-nll-02,bib-nlo-nll-03,bib-nlo-nll-04,bib-nlo-nll-05},
along with the ${\pm}1$~standard deviation
theory uncertainties~\cite{Kramer:2012bx}.
The dashed (red) contours present the corresponding expected results,
along with the ${\pm}1$~standard deviation experimental uncertainties.
}
\label{fig:observed-t1bbbb}
\end{figure*}

Upper limits on the cross sections to produce events in the
\tonebbbb and \tonetttt scenarios are determined at
95\% confidence level (CL).
The limits,
based on the CL$_\mathrm{s}$~\cite{Junk1999,bib-cls} technique
with the test statistic $q_\mu$ defined above,
are presented
as a function of the gluino and LSP masses.
Using the NLO+NLL cross section as a reference,
we also evaluate the corresponding 95\% CL exclusion curves.
The results are shown in Fig.~\ref{fig:observed-t1bbbb}.
The selection efficiency for \tonebbbb (\tonetttt) events
is fairly constant at about 60\% (25\%)
except for points to the left of a line parallel to the
diagonal that intersects the $\mlsp=0$ axis at around
$\mgluino=400\GeV$ (550\GeV)
or for gluino masses below about 550\GeV (680\GeV),
where the efficiency decreases smoothly to 15\% or less.
Conservatively using the minus-one-standard-deviation result~\cite{Kramer:2012bx}
for the reference cross sections,
and in the limit of a massless LSP,
we exclude gluinos with masses below 1170\GeV and 1020\GeV,
respectively,
in the \tonebbbb and \tonetttt scenarios.
While these limits do not exclude the entire range of gluino
masses $m_{\sGlu}\lesssim 1.5\TeV$ suggested by natural
models of SUSY~\cite{Papucci:2011wy},
they are nonetheless among the
most stringent bounds that have yet been obtained and greatly
improve our results from Ref.~\cite{RA2b2011pub}.

\section{Summary}
\label{sec:summary}

A search is presented for an anomalous rate of events
with three or more jets,
at least one bottom-quark-tagged jet,
no identified isolated electron or muon or isolated charged track,
and large missing transverse energy.
The search is based on a sample of proton-proton collision data
collected at $\sqrt{s}=8\TeV$ with the CMS detector at the LHC
in 2012,
corresponding to an integrated luminosity of 19.4\fbinv.
The principal standard model backgrounds,
from events with top quarks,
{\PW} bosons and jets,
{\Z} bosons and jets,
and QCD multijet production,
are evaluated using control samples in the data.
The analysis is performed in the framework of a global likelihood fit
in which the numbers of events in 165 exclusive bins
in a three-dimensional array of missing transverse energy,
the number of {\cPqb}-tagged jets,
and the scalar sum of jet \pt values,
are simultaneously examined.
The standard model background estimates are found to agree
with the observed numbers of events to within the uncertainties.
We interpret the results in the context of
simplified SUSY scenarios
in which gluino pair production is followed by the decay of
each gluino to an undetected particle and either a bottom or top quark-antiquark pair,
characteristic of gluino mediated bottom- or top-squark production.
Using the NLO+NLL production cross section as a reference,
and in the limit of a massless lightest supersymmetric particle,
we exclude gluinos with masses below 1170\GeV
and 1020\GeV for the two scenarios, respectively.
These are among the most stringent bounds that have yet been obtained
for gluino mediated bottom and top squark production.

\section*{Acknowledgements}

We congratulate our colleagues in the CERN accelerator departments for the excellent performance of the LHC and thank the technical and administrative staffs at CERN and at other CMS institutes for their contributions to the success of the CMS effort. In addition, we gratefully acknowledge the computing centres and personnel of the Worldwide LHC Computing Grid for delivering so effectively the computing infrastructure essential to our analyses. Finally, we acknowledge the enduring support for the construction and operation of the LHC and the CMS detector provided by the following funding agencies: BMWF and FWF (Austria); FNRS and FWO (Belgium); CNPq, CAPES, FAPERJ, and FAPESP (Brazil); MEYS (Bulgaria); CERN; CAS, MoST, and NSFC (China); COLCIENCIAS (Colombia); MSES (Croatia); RPF (Cyprus); MoER, SF0690030s09 and ERDF (Estonia); Academy of Finland, MEC, and HIP (Finland); CEA and CNRS/IN2P3 (France); BMBF, DFG, and HGF (Germany); GSRT (Greece); OTKA and NKTH (Hungary); DAE and DST (India); IPM (Iran); SFI (Ireland); INFN (Italy); NRF and WCU (Republic of Korea); LAS (Lithuania); CINVESTAV, CONACYT, SEP, and UASLP-FAI (Mexico); MSI (New Zealand); PAEC (Pakistan); MSHE and NSC (Poland); FCT (Portugal); JINR (Armenia, Belarus, Georgia, Ukraine, Uzbekistan); MON, RosAtom, RAS and RFBR (Russia); MSTD (Serbia); SEIDI and CPAN (Spain); Swiss Funding Agencies (Switzerland); NSC (Taipei); ThEPCenter, IPST and NSTDA (Thailand); TUBITAK and TAEK (Turkey); NASU (Ukraine); STFC (United Kingdom); DOE and NSF (USA).

Individuals have received support from the Marie-Curie programme and the European Research Council and EPLANET (European Union); the Leventis Foundation; the A. P. Sloan Foundation; the Alexander von Humboldt Foundation; the Belgian Federal Science Policy Office; the Fonds pour la Formation \`a la Recherche dans l'Industrie et dans l'Agriculture (FRIA-Belgium); the Agentschap voor Innovatie door Wetenschap en Technologie (IWT-Belgium); the Ministry of Education, Youth and Sports (MEYS) of Czech Republic; the Council of Science and Industrial Research, India; the Compagnia di San Paolo (Torino); the HOMING PLUS programme of Foundation for Polish Science, cofinanced by EU, Regional Development Fund; and the Thalis and Aristeia programmes cofinanced by EU-ESF and the Greek NSRF.

\bibliography{auto_generated}   

\cleardoublepage \appendix\section{The CMS Collaboration \label{app:collab}}\begin{sloppypar}\hyphenpenalty=5000\widowpenalty=500\clubpenalty=5000\textbf{Yerevan Physics Institute,  Yerevan,  Armenia}\\*[0pt]
S.~Chatrchyan, V.~Khachatryan, A.M.~Sirunyan, A.~Tumasyan
\vskip\cmsinstskip
\textbf{Institut f\"{u}r Hochenergiephysik der OeAW,  Wien,  Austria}\\*[0pt]
W.~Adam, T.~Bergauer, M.~Dragicevic, J.~Er\"{o}, C.~Fabjan\cmsAuthorMark{1}, M.~Friedl, R.~Fr\"{u}hwirth\cmsAuthorMark{1}, V.M.~Ghete, N.~H\"{o}rmann, J.~Hrubec, M.~Jeitler\cmsAuthorMark{1}, W.~Kiesenhofer, V.~Kn\"{u}nz, M.~Krammer\cmsAuthorMark{1}, I.~Kr\"{a}tschmer, D.~Liko, I.~Mikulec, D.~Rabady\cmsAuthorMark{2}, B.~Rahbaran, C.~Rohringer, H.~Rohringer, R.~Sch\"{o}fbeck, J.~Strauss, A.~Taurok, W.~Treberer-Treberspurg, W.~Waltenberger, C.-E.~Wulz\cmsAuthorMark{1}
% ...
\vskip\cmsinstskip
% ...
1:~~Also at Vienna University of Technology, Vienna, Austria\\
2:~~Also at CERN, European Organization for Nuclear Research, Geneva, Switzerland\\
% ...

\end{sloppypar}
\end{document}